\documentclass[10pt,dvipsnames,table, a4paper]{article}
\usepackage{setspace}
\usepackage{chngcntr}
\usepackage[authoryear,round]{natbib}
\counterwithout{equation}{section}

\usepackage[english]{babel}
\usepackage{multicol}
\usepackage{pifont}
\usepackage{graphicx}
\usepackage{amsthm}
\usepackage{amsmath}
\usepackage{booktabs}
\usepackage{amssymb}
\usepackage{mathtools}
\usepackage{bbm}
\usepackage{tabularx}
\usepackage[section]{placeins}
\usepackage{longtable}
\usepackage{colortbl}
\usepackage{afterpage}
\usepackage{wrapfig}
\usepackage{multirow}
\usepackage{pdflscape}
\usepackage{threeparttable}
\usepackage{threeparttablex}
\usepackage[resetlabels]{multibib}
\usepackage{lineno}
\usepackage{rotating}
\usepackage{titlesec}
\usepackage[
    labelfont=bf,
    font={footnotesize,singlespacing},
    singlelinecheck=false
]{caption}
\usepackage[framemethod=TikZ]{mdframed}
\usepackage[utf8]{inputenc}
\DeclareUnicodeCharacter{03B2}{\ensuremath{\beta}}
\usepackage[frozencache,cachedir=minted]{minted}
\usepackage{epigraph}
\usepackage{subcaption}
\usepackage{authblk}
\usepackage{makecell}

\usepackage{xcolor}
\newcounter{codeblock}
\renewcommand{\thecodeblock}{\arabic{codeblock}}

\newcommand{\cbref}[1]{Code Block~\ref{#1}}
\newcommand{\suppcbref}[1]{Supplementary Code Block~\ref{#1}}

\newenvironment{codeblock}[2]{%
  \refstepcounter{codeblock}\label{#2}%
  \begin{myenv}{Code Block \color{darkred}\thecodeblock\color{black}: #1}%
}{%
  \end{myenv}%
}

\newenvironment{suppcodeblock}[2]{%
  \refstepcounter{codeblock}\label{#2}%
  \begin{myenv}{Supplementary Code Block \color{darkred}\thecodeblock\color{black}: #1}%
}{%
  \end{myenv}%
}

\definecolor{darkblue}{RGB}{0,33,71}
\definecolor{darkred}{RGB}{153, 70, 54}
\usepackage[colorlinks=true,
linkcolor=darkred,
urlcolor=darkblue,
citecolor=darkblue]{hyperref}%
\usepackage[top=1.5cm,right=2cm,left=2cm,bottom=2cm]{geometry}
\usepackage{amsfonts}

\newcites{appendix}{Supplemental References}

\newcounter{subsubsubsection}[subsubsection]
\renewcommand\thesubsubsubsection{\thesubsubsection.\arabic{subsubsubsection}}
\titleclass{\subsubsubsection}{straight}[\subsubsection]
\titleformat{\subsubsubsection}[block]
  {\normalfont\normalsize\bfseries}{\thesubsubsubsection\quad}{0pt}{}
\titlespacing*{\subsubsubsection}{0pt}{3.25ex plus 1ex minus .2ex}{1.5ex plus .2ex}
\newenvironment{myenv}[1]
  {\mdfsetup{
    frametitle={\colorbox{white}{\space#1\space}},
    innertopmargin=0pt,
    frametitleaboveskip=-\ht\strutbox,
    frametitlealignment=\center,
    linewidth=1pt,
    roundcorner=10pt
    }
  \begin{mdframed}%
  }
  {\end{mdframed}}

\makeatletter
\def\maxwidth{\ifdim\Gin@nat@width>\linewidth\linewidth\else\Gin@nat@width\fi}
\def\maxheight{\ifdim\Gin@nat@height>\textheight\textheight\else\Gin@nat@height\fi}

\makeatother

\newcommand\blfootnote[1]{%
  \begingroup
  \renewcommand\thefootnote{}\footnote{#1}%
  \addtocounter{footnote}{-1}%
  \endgroup
}

\makeatletter
\renewcommand{\thanks}[1]{%
  \begingroup
  \renewcommand\thefootnote{}%
  \footnotetext{#1}%
  \endgroup
}
\makeatother

\newcommand{\supplementarysection}{%
  \setcounter{figure}{0}% Reset figure counter
  \let\oldthefigure\thefigure% Capture figure numbering scheme
  \renewcommand{\thefigure}{S\oldthefigure}% Prefix figure number with S
  \section{Supplementary section}% Set supplementary section
}

\title{\Large{RobustiPy: An efficient next-generation multiversal library with model selection, averaging, resampling, and explainable AI}

\blfootnote{For correspondence: Daniel Valdenegro Ibarra, Jiani Yan, and Charles Rahal, Leverhulme Centre for Demographic Science and the Centre for Care, Oxford Population Health, University of Oxford, OX1 1JD, United Kingdom. Tel: 01865 286170. Email: \href{mailto:daniel.valdenegro@demography.ox.ac.uk}{daniel.valdenegro@demography.ox.ac.uk}, \href{mailto:jiani.yan@demography.ox.ac.uk}{jiani.yan@sociology.ox.ac.uk} and \href{mailto:charles.rahal@demography.ox.ac.uk}{charles.rahal@demography.ox.ac.uk}. The authors declare no conflicts of interest and are grateful to comments on earlier versions of the work from participants at the International Conference on Computational Social Science, the British Society for Population Studies, the European Consortium for Sociological Research, Population Association of America, Tsinghua University, the International Conference on Social Computing, and internal feedback received from the Leverhulme Centre for Demographic Science, Reproducible Research Oxford, and the Pandemic Sciences Institute. We also thank all attendees at a previous hackathon, hosted at the University of Oxford in June 2024. We especially thank Anda-Raluca Epure for detailed comments on the manuscript and an extensive code review of the accompanying software. The authors are grateful for funding from the ESRC (an ESRC Large Centre grant), the Leverhulme Trust (Grant RC-2018-003) for the Leverhulme Centre for Demographic Science and Nuffield College, and a Grand Union DTP ESRC studentship.
}}

\author[1,3,5]{\normalsize{Daniel Valdenegro Ibarra}}
\author[1,2,3,4]{\normalsize{Jiani Yan}}
\author[1,3]{\normalsize{Duiyi Dai}}
\author[1,3,5]{\normalsize{Charles Rahal}}

\affil[1]{Leverhulme Centre for Demographic Science, University of Oxford}
\affil[2]{Department of Sociology, University of Oxford}
\affil[3]{ESRC Centre for Care, University of Oxford}
\affil[4]{Wolfson College, University of Oxford}
\affil[5]{Nuffield College, University of Oxford}

\AtBeginDocument{%
  \counterwithout{equation}{section}%  stop resetting by section
}


\begin{document}

\maketitle
\vspace{-0.25in}
\begin{abstract}
	Scientific inference is often undermined by the vast but rarely explored `multiverse' of defensible modeling choices which can generate results as variable as the phenomena under study. We introduce \texttt{RobustiPy}, an open-source Python library that systematizes multiverse analysis and model-uncertainty quantification at scale. \texttt{RobustiPy} unifies bootstrap-based inference, combinatorial specification search, model selection and averaging, joint-inference routines, and explainable AI methods within a modular, reproducible framework. Beyond exhaustive specification curves, it supports rigorous out-of-sample validation and quantifies the marginal contribution of each covariate. We demonstrate its utility across five simulation designs and ten empirical and high-profile replications spanning economics, sociology, psychology, and medicine, including a re-analysis of widely cited findings with documented discrepancies. Benchmarking on $\sim$672 million simulated regressions shows that \texttt{RobustiPy} delivers state-of-the-art computational efficiency while expanding transparency in empirical research. By standardizing and accelerating methods for robustness, \texttt{RobustiPy} transforms how researchers interrogate sensitivity across the analytical multiverse, offering a practical foundation for more reproducible and interpretable computational science.
\end{abstract} \vspace{.05in}

\textit{Open Science}

\newpage

\setlength{\epigraphwidth}{0.62\textwidth}
\epigraph{``\textit{Analysts can test thousands of plausible models, trying out different model assumptions. Consumers are still constrained to know only the handful of model specifications that are curated for publication. Technology created this problem, and technology will have to be a part of the solution}.''} {-- \cite{young2025multiverse}} \vspace{.05in}

\section{Introduction}

The act of constructing quantitative models to examine scientific phenomena represents the cornerstone of most intellectual processes \citep{oreskes2003role}. Yet, the scientific community -- particularly within the health and social sciences \citep{camerer2018evaluating} -- has long recognized that the process of testing hypotheses using such models is susceptible to substantial variation in outcomes, even under relatively conservative assumptions \citep{klein2014investigating}. While Leamer's infamous “Let’s Take the Con Out of Econometrics” article \citep{leamer1983let} is often credited with dramatically drawing attention to pervasive risks in applied work, the concept and even formal tools for handling model uncertainty date back decades earlier \citep{jeffreys1939theory,lindley1957statistical}. Numerous factors influence the model-building process, ranging from uncontrollable elements to more structured components like the conditions of data collection; the choices which responsible researchers make over this space are otherwise known as the `garden of forking paths' \citep{gelman2013garden}. The range of possible outcomes in scientific modeling may, in fact, mirror the vast variation inherent in reality itself, all when focusing on just one estimand of interest \citep{lundberg2021your}. At the same time, it is increasingly recognized that complex computational routines can themselves be utilised for theory-building \citep{engzell2023understanding}. We create a tractable solution which takes steps to reduce this consequential source of (epistemic\citealp{yan2025unknowable}) variation; the analytical choices made by researchers. These choices can -- in principle -- be systematically examined and controlled. Developed in \texttt{Python}, our readily available, continuously maintained, and highly accessible tool aims to bring transparency to research practice, representing what we believe is the next-generation of model-uncertainty software. \texttt{RobustiPy} enables efficient large-scale specification curve and multiverse analyses within a unified, reproducible ecosystem; it is also agnostic to philosophies regarding whether models in a multiverse should be weighted \citep{young2019difference, slez2019difference} by considering both the full model space, metrics of fit, estimates which maximize information criteria, and weighted and unweighted estimates of the estimand. We document the five different types of analysis available through simulated analysis, conduct a time-profiling exercise, and apply \texttt{RobustiPy} to ten impactful empirical examples across the research and teaching-related landscape. These empirical examples span the economic, social, psychological and medical sciences, and beyond. All of our simulated and empirical examples are made available as \texttt{Python} scripts and interactive notebooks within a release of our library which has been indexed into Zenodo \citep{valdenegro_ibarra_2025_15700698}.\par

Analytical decisions -- in comparison to more ethereal, uncontrollable concerns -- are not only more amenable to scrutiny, but also potentially more influential for the scientific enterprise. While other sources of variation (such as data collection) may be expected to cancel out with replication, research decisions regarding the modeling of scientific processes are often inherently non-random and systematically biasing. They can introduce systematic biases that remain hidden unless explicitly interrogated. This typically results in an asymmetry of information \citep{young2025multiverse}: current common practice involves researchers varying their decisions, selectively reporting what they choose. Although exploring alternative specifications is not inherently problematic, the persistent lack of transparency in these practices enables questionable research behaviours, such as p-hacking; the process of iteratively modifying models until statistically significant results are obtained \citep{head2015extent, brodeur2016}. A related issue is `HARKing' (Hypothesising After the Results are Known, see for example \citealp{kerr1998harking}), in which researchers test multiple hypotheses and present significant findings as though they were specified in advance. Beyond concerns about scientific integrity, such selective reporting contributes to poor reproducibility and a misleading sense of empirical consensus, ultimately undermining confidence in research and the scientific endeavour more broadly \citep{wingen2020no}. A range of solutions have been proposed to address these challenges. These include greater transparency through preregistration \citep{van2016pre}, a stronger emphasis on replication \citep{edlund2022saving}, improved statistical power \citep{head2015extent, benjamin2018redefine}, the adoption of Bayesian inference \citep{etz2016bayesian}, and a shift towards predictive modeling as an evaluation tool \citep{verhagen2024incorporating, rahal2022rise}. One particularly promising approach to improving transparency is the systematic reporting of all defensible combinations of analytical decisions made by a researcher. This strategy -- independently developed as specification curve analysis \citep{simonsohn2020specification} and multiverse analysis \citep{steegen2016increasing} -- explicitly counters selective reporting; the former by looking at the specification-based choices which researchers can make, and the latter by analysing a broader choice space. Rather than presenting a single preferred model, it aims to identify and estimate all plausible analysis branches, presenting the full distribution of permissible results in a single, usually visual output. This is what we build upon.

Despite their conceptual appeal, multiverse and specification-curve analyses are computationally demanding. Among the many degrees of freedom available to researchers, covariate selection alone can result in an exponential number of model specifications. For example, a dataset with ten candidate control variables produces $2^{10} = 1,024$ possible models; twenty control variables yield over a million. Historically, researchers have tackled this complexity using bespoke computational routines to estimate and summarise the results of thousands \citep{leamer1983let}, millions \citep{sala1997just}, billions \citep{munoz2018we}, or even trillions \citep{hanck2016just} of regression models. The expansion of empirical research has brought increased attention to the role of `researcher degrees of freedom' -- the wide range of defensible yet subjective decisions made during data collection, model specification and result reporting \citep{gelman2014statistical}. All of which undermine the validity of empirical findings and contribute to broader issues such as the replication crisis. These problems are pervasive across almost all quantitative academic disciplines, spanning and especially psychology \citep{simmons2011false}, the social sciences \citep{huntington2021influence}, and medical research \citep{ioannidis2005most}. However, custom implementations are often inefficient and insufficiently feature-rich, despite the ever increasing computational power available to researchers in the limit of human progress \citep{yan2025unknowable, sutton2019bitter}. To overcome these limitations, we introduce \texttt{RobustiPy}, which streamlines model selection, averaging, and out-of-sample validation and feature contribution while performing statistical inference testing and evaluating fit across the entire space of defensible models. By offering a standardised, efficient, and unit-tested solution, \texttt{RobustiPy} enhances computational reproducibility and mitigates the risks associated with costly and limited ad-hoc implementations of multiverse and specification curve analysis.

\subsection{Existing State-of-the-Art}

Several computational tools have been developed to facilitate the implementation of specification curve analysis -- a way to visualise all specifications of a model space -- and multiverse analysis -- the systematic estimation of every defensible analytic pathway -- in applied research. In the \texttt{R} programming ecosystem, the \textit{multiverse} package by \citet{sarma2023multiverse} introduces a declarative syntax for defining multiple decision points within a single analysis script. Users can specify branches corresponding to alternative models and specifications, which are then executed across the entire, permissible model space; the package integrates with \texttt{R} Markdown and supports structured workflows for running and reporting multiverse analyses. Specifically regarding specification curve analysis -- as opposed to multiversal analysis -- and likely most formally introduced by \cite{simonsohn2020specification}, the \textit{specr} package \citep{masur2020specr} offers a comprehensive set of tools to define, estimate, and visualise a range of model specifications. It is particularly well suited for observational studies, enabling researchers to explore variations in outcome variables, predictor variables, and control variables adjustments. Results are presented through specification curves that summarise the distribution of estimates and confidence intervals across the analytical space. \cite{semken2022specification} advocate a Bayesian approach, with code available on the Open Science Framework \citep{semken2022supplemental}. Similar, but with more limited functionality, \emph{spec\_curve} allows you to create a specification chart using base \texttt{R} \citep{OrtizBobea2021spec_chart}. In \texttt{Stata}, there exists the \emph{MULTIVRS} module to conduct multiverse analysis \citep{RePEc:boc:bocode:s458927}, and \emph{MROBUST}, a module to estimate model robustness and model influence \citep{young2017model}.

In contrast, the development of comparable toolkits in the \texttt{Python} ecosystem has been more limited, despite the fact that \texttt{Python} has become the world's most widely used programming language for highly performant data science research \citep{castro2023landscape}. The \textit{spec\_curve} package by \citet{aeturrell_specification_curve_2025} provides foundational functionality for conducting specification curve analyses. While it offers a useful entry point, it lacks the breadth and extensibility of its \texttt{R}-based counterparts. The Boba domain-specific language (DSL, see \citealp{liu2020boba}) provides a programming language agnostic tool and visualisation system for multiverse analysis. It automatically expands and executes all valid combinations of user decision points. The results are then presented through an interactive, linked-view interface. While Boba supports execution in both \texttt{R} and \texttt{Python} environments, its use of a custom DSL makes it difficult to integrate seamlessly into standard analysis pipelines. \texttt{RobustiPy} extends beyond existing tools by incorporating functionality for model selection, bootstrapping, joint inference, out-of-sample evaluation, and the systematic exploration of multiple estimands and outcome spaces. In doing so, it enables a more rigorous and transparent approach to robustness analysis. A summary of the main packages in this space is described in Table \ref{tab:table1}.

\begin{table}[!t]
	\small
	\renewcommand{\arraystretch}{1.25} % Adds extra space between rows
	\centering
	\caption{Comparison of Toolkits for Multiverse and Specification Curve Analysis}
	\begin{tabularx}{\textwidth}{lXXXXXX}
		\toprule
		\textbf{Feature / Toolkit}
		                                 & \textbf{\texttt{multiverse}\newline(R)}
		                                 & \textbf{\texttt{specr}\newline(R)}
		                                 & \textbf{\texttt{MULTIVRS}\newline(Stata)}
		                                 & \textbf{\texttt{spec\_curve}\newline(Python)}
		                                 & \textbf{\texttt{Boba}\newline\scriptsize{(Python based DSL)}}
		                                 & \textbf{\texttt{RobustiPy}\newline(Python)}
		\\
		\midrule
		Multiverse analysis\footnotemark & \checkmark                                                    & --                   & \checkmark\footnotesize{} & --                  & \checkmark           & \checkmark (limited) \\
		Specification curve analysis     & --                                                            & \checkmark           & --                        & \checkmark          & --                   & \checkmark           \\
		High-abstraction syntax          & \checkmark                                                    & \checkmark (partial) & \textemdash (limited)     & --                  & \checkmark           & \checkmark           \\
		Visualisation of results         & \checkmark                                                    & \checkmark(built-in) & \checkmark (kdensity)     & \checkmark(basic)   & \checkmark           & \checkmark (custom)  \\
		Joint inference                  & --                                                            & \checkmark           & --                        & \checkmark          & \checkmark           & \checkmark           \\
		Influence analysis               & --                                                            & --                   & \checkmark\footnotesize{} & --                  & \checkmark           & \checkmark           \\
		Bootstrapping                    & --                                                            & \checkmark           & --                        & \checkmark          & \checkmark (partial)
		                                 & \checkmark                                                                                                                                                                           \\
		Out-of-sample                    & --                                                            & --                   & --                        & --                  & --                   & \checkmark           \\
		Multiple estimands               & --                                                            & \checkmark (limited) & --                        & \checkmark(partial) & --                   & \checkmark           \\
		\bottomrule
	\end{tabularx}
	\label{tab:table1}
\end{table}

\footnotetext{In this table, we consider ``multiverse" analysis the analysis that is able to provide outputs using different estimators (e.g. linear regression, logistic regression, probit regression, etc.)}

\subsection{Formalisation}\label{sec:formalisation}

In order to describe the problem which \texttt{RobustiPy} addresses, we formalise the problem below. % To do this we will borrow the notation from \cite{simonsohn2020specification}.
Consider the key association between variables $\mathbf{Y}$ and $\mathbf{X}$, where a set of control variables $\mathbf{Z}$ can influence the relationship between the former as follows:

\begin{equation}\label{objective_func}
	\mathbf{Y} = F(\mathbf{X}, \mathbf{Z}) + \boldsymbol{\epsilon} .
\end{equation}

\noindent Assume $\mathbf{Y}$ is an unknown data generating function $F()$ which takes the values of $\mathbf{X}$ conditioned by the values of the set of control variables $\mathbf{Z}$, plus a residual term ($\boldsymbol{\epsilon}$). In this conceptualisation of the problem, Eq. \ref{objective_func} represents the real data generation process that researchers approximate, but, its perfect reproduction is (almost always) unattainable. Following the example of \cite{simonsohn2020specification}, $\mathbf{Y}$ and $\mathbf{X}$ are usually imprecisely defined latent variables (for which we cannot have a direct measure). Likewise, the set of control variables $\mathbf{Z}$ can also be composed of imprecisely defined variables; the number of them can be unknown and/or non-finite. Under these conditions, it is evident that researchers must make concessions, usually in the form of finite sets of `reasonable' operationalisations of $\mathbf{Y}$, $\mathbf{X}$, $\mathbf{Z}$ and $F()$. The way `reasonable' is defined is in itself a point of contention; \cite{simonsohn2020specification} somewhat canonically defines this as a set where each element is theoretically justified, statistically valid, and non-redundant (we prefer the term `defensible'). Let's define the set of `defensible' operationalisations of $\mathbf{Y}$, $\mathbf{X}$, $\mathbf{Z}$ and $F()$ as $\overleftrightarrow{\mathbf{Y}}$, $\overleftrightarrow{\mathbf{X}}$, $\overleftrightarrow{\mathbf{Z}}$ and $\overleftrightarrow{F()}$. Then, we have:

\begin{equation}\label{spec_func}    \overleftrightarrow{\mathbf{Y}}_{\pi} = \overleftrightarrow{F}_{\pi}(\overleftrightarrow{\mathbf{X}}_{\pi}, \overleftrightarrow{\mathbf{Z}}_{\pi}) + \boldsymbol\epsilon.
\end{equation}

\noindent Eq. \ref{spec_func} corresponds to a single possible choice ($\pi$) of all the possible choices, $\Pi$ -- such that $\pi \in \Pi$ -- which we can generate using our current formalisation. The total number of estimations is the number of distinct combinations obtained by selecting (i) one composite outcome operationalisation (potentially created by one or many components), (ii) one functional form (estimator), (iii) one focal predictor specification, and (iv) a subset of candidate controls.
Let
\(
m_F \coloneqq |\Pi_{\overleftrightarrow{F()}}|,
\;
m_X \coloneqq |\Pi_{\overleftrightarrow{\mathbf{X}}}|.
\)
If there are \(d_Z\) candidate control variables and \(d_Y\) candidate components of a composite dependant variable, and the modelling decision is inclusion/exclusion of each control and candidate for dependant composition (with a minimum of at least one candidate), then
\(
|\Pi_{\overleftrightarrow{\mathbf{Z}}}| = 2^{d_Z}
\)
and \(
|\Pi_{\overleftrightarrow{\mathbf{Y}}}| = 2^{d_Y}-1
\)
.
Hence:
\begin{equation}\label{eq:Pi_size}
	|\Pi|
	= (2^{d_Y}-1)\,m_F\,m_X\,2^{d_Z}.
\end{equation}
For example, with two functional forms (\(m_F=2\)), two target variables (\(d_Y=2\)),
two alternative focal predictors (\(m_X=2\)), and four candidate controls (\(d_Z=4\)),
we have \(|\Pi| = (2^2 -1) \times 2 \times 2 \times 2^{4} = 192\).

%It is unlikely that the computation of the specification space $\Pi$ would be done whole in one single step. It is far more practical and clear to split the computation into its main components, so resources can be focused into the more demanding tasks.
We split the computation based on the operationalised variables in Eq. \ref{spec_func} by creating a `defensible specification space' for each: $\Pi_{\overleftrightarrow{\mathbf{Y}}}$, $\Pi_{\overleftrightarrow{F()}}$, $\Pi_{\overleftrightarrow{\mathbf{X}}}$, $\Pi_{\overleftrightarrow{\mathbf{Z}}}$. The entire independent model space can be expressed as:

\begin{equation}
	\Pi = \Pi_{\overleftrightarrow{\mathbf{Y}}} \times \Pi_{\overleftrightarrow{F()}} \times \Pi_{\overleftrightarrow{\mathbf{X}}} \times \Pi_{\overleftrightarrow{\mathbf{Z}}}
\end{equation}

\noindent Given that $\Pi$ represents a collection of multiple possible versions of the true relationship between $\mathbf{Y}$ and $\mathbf{X}$, an increasingly large space of random possible choices ($D$) will in principle always lead to a more reasonable approximation of $\mathbf{Y} = F(\mathbf{X},\mathbf{Z})$. The main problem with current research practices is that researchers usually choose a small and non-random sample of ${\Pi}$ to represent the variety of results in the full space of possible choices \citep{simonsohn2020specification}.\footnote{This space is usually as sparse as a single `baseline' model and a few scant robustness checks -- which coincidentally always seem to concur with the results from the baseline -- at the request of peer reviewers; see  \cite{young2017model} for a quantification of this at two leading Sociology journals.} %Since:
%\begin{equation}
%    \lim_{D \to \infty}\Pr \left( \mathbb{E}[\overleftrightarrow{\mathbf{Y}}_{D}] = \mathbf{Y} \right) = 1,
%\end{equation}

%\begin{equation}
%    \lim_{D \to \infty} \mathop{E}[\overleftrightarrow{Y}%_{\Pi}] - \mathop{E}[\overleftrightarrow{Y}^{D}_{\hat{\Pi}}] = 0.
%\end{equation}

%\noindent and as a corollary, we propose that researchers should strive to consider as large a combinatorial set of different model instantiations ($\Pi$) as possible in order to have a better chance of approximating the real data generation process. %If such amount of operationalisations leads to a computationally intractable set of model specifications, we can still have a good approximation of the real data generation problem by randomly sampling such specification space. \par

We formalise the principle that enlarging the catalogue of admissible models improves approximation to the truth.
Let $\{\Pi_D : D \in \mathbb N\}$ be a nested sequence of specification sets with $\Pi_D \subset \Pi_{D+1}$ and cardinality $N_D \coloneqq |\Pi_D| \to \infty$ as $D\to\infty$, where $D$ indexes model complexity. For each $\pi \in \Pi_D$, let $\widehat Y_{\pi}(X,Z)$ denote the fitted prediction based on a training sample $\mathcal S_n$, and define the corresponding mean predictor
\[
	\bar Y_{\pi}(x,z) := \mathbb E_{\mathcal S_n}\!\left[\,\widehat Y_{\pi}(x,z)\,\right].
\]
The target regression function is $F^{\star}(x,z) = \mathbb E[Y \mid X=x, Z=z]$.
All $L_2$ norms are taken with respect to the distribution $P_{X,Z}$ of the covariates:
\[
	\|g\|_{L_2(P)} = \left(\int g(x,z)^2\,\mathrm dP_{X,Z}(x,z)\right)^{1/2}.
\]

\noindent Since $\Pi_D$ may be too large to explore exhaustively, we allow for a random sub-selection $\mathcal R_D \subseteq \Pi_D$.
Assume (i) the union $\bigcup_D \Pi_D$ is dense in $L_2(P)$, and (ii) there exists $c>0$ such that for all $D$ and all $\pi \in \Pi_D$,
\[
	\Pr\!\left(\pi \in \mathcal R_D\right) \;\geq\; c.
\]

\noindent Then, for every $\varepsilon > 0$,
\begin{equation}
	\label{eq:oracle-approx}
	\lim_{D \to \infty}
	\Pr_{\mathcal R_D}\!\Biggl[
	\inf_{\pi \in \mathcal R_D}
	\bigl\| \bar Y_{\pi} - F^\star \bigr\|_{L_2(P)} < \varepsilon
	\Biggr]
	= 1 .
\end{equation}

\noindent This establishes an oracle approximation property (cf.\ \citealp{barron1999risk, Chen2007Sieve}).

\subsection{Types of Analysis}

\texttt{RobustiPy} enables an approximation of the real data generation process (Eq. \ref{objective_func}) by generating and organising specifications spaces ($\Pi_{\overleftrightarrow{\mathbf{Y}}}$, $\Pi_{\overleftrightarrow{\mathbf{X}}}$, $\Pi_{\overleftrightarrow{\mathbf{Z}}}$). In particular, \texttt{RobustiPy} explicitly encourages a full examination across the entirety of  $\Pi_{\overleftrightarrow{\mathbf{Z}}}$ assuming a single $\overleftrightarrow{\mathbf{Y}}$, a fixed set of $\overleftrightarrow{\mathbf{X}}$ predictors and a fixed set of $\overleftrightarrow{F()}$ estimators.\footnote{We call this the `Vanilla' computation in Sections \ref{sec_types:vanilla} and \ref{sec:results_vanilla}.} Additional functionality allows users to generate and compute $\Pi_{\overleftrightarrow{\mathbf{Y}}}$, assuming multiple valid dependent measurements. Let's review the most common use cases -- including two stand-alone Code Blocks -- where our online repository hosts five simulations pertaining to each type of analysis and Section \ref{emp_and_sim} discusses ten highly varied empirical applications of each of the five types. Section \ref{discussion} discusses these results, Section \ref{availability} details code and data availability, and Section \ref{sec:online_methods} contains a full delineation of every metric and method which is undertaken and available within \texttt{RobustiPy}.

\subsubsection{Vanilla computation}\label{sec_types:vanilla}

Our first simulated example (\cbref{cb:vanilla}) assumes a linear data generating process. Here we fit an OLS estimator with 10‐fold cross‐validation and 1,000 bootstrap resamples, and then we plot results and print a large number of summary outputs. The data generation process described in \cbref{cb:vanilla} is analogous to a typical modeling problem whereby a researcher seeks to evaluate the relationship between a target variable of interest with only one valid operationalisation, and a predictor with only one possible operationalisation which is conditional on a `reasonable' set of control variables (and, by extension, a researcher's idea about the data generation process, be it theoretical or otherwise). In this situation (i.e. with no variation in the outcome space, the variable of interest space, or the functional form), $\Pi_{\overleftrightarrow{\mathbf{Z}}} \equiv \Pi$. For the purposes of delineation -- as in \cbref{cb:vanilla} -- the \texttt{OLSRobust} class allows us to declare an object (\texttt{vanilla\_obj}) containing the target variable ($Y_{1}$), the main predictor as part of an array ($\overleftrightarrow{\mathbf{X}}$), and the source of data. The \texttt{.fit()} method allows us to declare the array of possible control variables ($\overleftrightarrow{\mathbf{Z}}$) -- which a researcher may be agnostic about including -- to be used in the analysis. \texttt{RobustiPy} automatically adds an intercept unless the user has already included one in $\overleftrightarrow{\mathbf{X}}$ or $\overleftrightarrow{\mathbf{Z}}$ (i.e., a feature with any name which has zero variance). When a constant is present in $\overleftrightarrow{\mathbf{X}}$, the software leaves the matrix unchanged. A constant placed in $\overleftrightarrow{\mathbf{Z}}$ is treated as a varying regressor (with any constants in $\overleftrightarrow{\mathbf{X}}$ removed) such that the iterator yields some specifications with an intercept, and others without. We -- by default -- parallelise computation to decrease runtime.\footnote{\texttt{RobustiPy} supports interactive prompting for \texttt{n\_cpu}. In interactive sessions, if \texttt{n\_cpu} is not supplied, the user can confirm or choose the CPU count; in non-interactive contexts, defaults are applied.} Finally,\ to return the results of fitted models, \texttt{OLSRobust} has a method entitled \texttt{get\_results}, returning a class object containing estimates as well as utility functions for summarisation and visualisation. We allow the user to specify which specifications they wish to visually highlight -- allowing direct comparisons with already published results -- alongside multiple other options.\footnote{By default, the `full' model (the one containing all $\overleftrightarrow{\mathbf{Z}}$ control variables) and the `null' model (which contains only the smallest guaranteed model space) are always highlighted in the relevant visual analysis for the case of one dependent variable. When there is more than one dependent variable specified -- see Section \ref{sec1:multiple_y} -- it highlights the model with solely the first dependent variable specified in the list, and the model which incorporates all dependent and independent variables.}

\begin{figure}[!t]
	\begin{codeblock}{Vanilla computation}{cb:vanilla}
		\begin{minted}[baselinestretch=1]{python}
import numpy as np
import pandas as pd
from robustipy.models import OLSRobust


def sim1(project_name):
    # 1. Setup
    n = 100
    rng = np.random.default_rng(192735)
    z = [f"z{i}" for i in range(1, 5)]

    # 2. Simulate covariates
    X = rng.standard_normal((n, 1))
    Z = rng.standard_normal((n, len(z)))
    df = pd.DataFrame(np.hstack([X, Z]), columns=["x1"] + z)

    # 3. Generate outcome y1
    df["y1"] = (1 + 2 * df["x1"] + 0.5 * df[z].sum(axis=1)
                + rng.standard_normal(n) * 0.5)

    # 4. Fit robust OLS with cross‐validation
    vanilla_obj = OLSRobust(y=["y1"], x=["x1"], data=df)
    vanilla_obj.fit(controls=z, draws=1000, kfold=10, seed=192735)

    # 5. Retrieve and plot results
    vanilla_res = vanilla_obj.get_results()
    vanilla_res.plot(specs=[z[1:3]], figsize=(16, 16), ci=0.95,
                     ic="bic", ext="svg", figpath='../figures',
                     project_name=project_name)
    vanilla_res.summary()


if __name__ == "__main__":
    sim1('sim1_example')
\end{minted}
	\end{codeblock}
\end{figure}

\subsubsection{Array of never changing variables}\label{sec1:never_changing}

\texttt{RobustiPy} also allows for the inclusion of multiple `fixed’ predictors, which are not varied across the covariate space. Programmatically, this set of fixed predictors will be included in all specifications, but not considered in the combinatoric operation. They act as a set of always present features. This functionality is useful in cases when the researchers have a strong determination that these variables always need to be included in a model, likely due to theoretical reasoning (or for specific hypothesis testing, see Section \ref{sec:results_constants}). It also has the secondary effect of reducing the specification space and hence the compute cost, making it an alternative path to follow in cases with very large sets of covariates (see Section \ref{results:fixed_effects}). To do this, we simply need to append additional predictors into our $\overleftrightarrow{\mathbf{X}}$ array; \texttt{RobustiPy} will take the first feature name of the list as the main estimand of interest, and the rest as fixed, never varying predictors. A stand-alone example of this type of use case is shown in \suppcbref{scb:never-changing}. An alternate approach to reduce compute cost with large control variables space is to pass \texttt{z\_specs\_sample\_size=} in \texttt{OLSRobust.fit(...)}, which samples the covariate space.

\subsubsection{Fixed effect OLS}

\texttt{RobustiPy} also provides the option to run a fixed effects OLS model. These types of models are commonly used in the analysis of panel data, in which repeated observations are nested within the observed unit (e.g. individuals, households, etc.), usually over time. In these contexts, fixed effects OLS is used to isolate and eliminate the intraclass variation of the observations. \texttt{RobustiPy} allows the specifying of a grouping variable, `\texttt{group}', which enters into the \texttt{.fit()} method. By providing this, \texttt{RobustiPy} automatically performs demeaning of the relevant variables and estimates its sets of models on the resultant, transformed data. In the fixed effect model, the default constant will be removed before any model-fitting as it contains zero information after demeaning.\footnote{This is because a constant column will be `1' for all rows and becomes all `0's after demeaning. If there are group-invariant variables,  \texttt{RobustiPy} will send warning messages about their existence but not remove them from the model.} Grouping identifiers may be string/categorical; numeric encoding is not required. A stand-alone example use case is shown in \suppcbref{scb:fixed}.

\subsubsection{Binary dependents}
\texttt{RobustiPy} also provides the opportunity to apply a logistic estimator to handle binary outcome variables (via the Newton method, see \citealp{nelder1972generalized}, and a tolerance of $1\times10^{-7}$). We chose logistic regression as it is one of the most common and computationally efficient estimators for binary choice models; an important consideration when running specification curves. This functionality is implemented through the \texttt{LRobust} class, which offers essentially the same functionality as \texttt{OLSRobust}, but only accepts binary outcome variables. When \texttt{group} is supplied to \texttt{LRobust}, it is used for grouped cross-validation and grouped bootstrap resampling, not for within-group demeaning. Naturally, it is also possible to estimate the model space for a binary dependent variable with \texttt{OLSRobust}, which would represent a Linear Probability Model (see, for example \citealp{aldrich1984linear}). A stand-alone example use case is shown in \suppcbref{scb:logisticregression}.

\begin{figure}[!t]
	\begin{codeblock}{Multiple dependents}{cb:multipledeps}
		\begin{minted}[baselinestretch=1]{python}

import numpy as np
import pandas as pd
from robustipy.models import OLSRobust


def sim5(project_name):
    # 1. Simulation parameters
    n, p = 1000, 5
    sigma_X = np.eye(p)
    rng = np.random.default_rng(192735)

    # 2. Draw covariates X and error epsilon
    X = rng.multivariate_normal(mean=np.zeros(p), cov=sigma_X, size=n)
    epsilon = rng.standard_normal(n)

    # 3. Stack β‐vectors into a (4xp) matrix
    B = np.array([
        [0.20, 0.50, -0.40, -0.10, 0.20], [0.30, 0.40, -0.35, -0.10, 0.20],
        [0.15, 0.60, -0.45, -0.10, 0.20], [0.40, 0.30, -0.50, -0.10, 0.20],
    ])  # shape = (4, p)

    # 4. Build DataFrame/generate outcomes and controls
    df = pd.DataFrame(X, columns=[f'x{i}' for i in range(1, p + 1)])
    df[[f'y{j}' for j in range(1, 5)]] = X @ B.T + epsilon[:, None]
    df[[f'z{j}' for j in range(1, 5)]] = rng.standard_normal((n, 4))

    # 5. Specify outcome and control names
    Y = [f'y{j}' for j in range(1, 5)]
    z = [f'z{j}' for j in range(1, 5)]

    # 6. Fit robust OLS
    model = OLSRobust(y=Y, x=['x1'], data=df)
    model.fit(controls=z, draws=1000, kfold=10, rescale_y=True,
              rescale_x=True, rescale_z=True, seed=192735)

    # 7. Retrieve and plot results
    res = model.get_results()
    res.plot(loess=False,
        specs=[['y1', 'y2', 'z1', 'z2'], ['y4', 'z4']],
        figsize=(16, 8), figpath='../figures', project_name=project_name)
    res.summary()

if __name__ == "__main__":
    sim5('sim5_example')
\end{minted}
	\end{codeblock}
\end{figure}

\subsubsection{Multiple dependents}\label{sec1:multiple_y}
Let's now modify our assumed data generation process by allowing the dependent variable to be a vector of four possible operationalisations. We describe the process to simulate this data in \cbref{cb:multipledeps}. In this scenario, researchers might not only want to compute $\Pi_{\overleftrightarrow{\mathbf{Z}}}$ over each operationalisation in $\overleftrightarrow{\mathbf{Y}}$ but also to compute all the possible combinations of composites over $\overleftrightarrow{\mathbf{Y}}$ (or, in other words, $\Pi_{\overleftrightarrow{\mathbf{Y}}}$) in order to properly obtain $\Pi$. It might also involve an array of never-changing predictors (Section \ref{sec1:never_changing}). \texttt{RobustiPy} takes the list of operationalisations and creates combinations of composites by first standardising their values, and then combining them using the row-wise average. Assume $m$ alternative operationalisations of $Y = \{Y^{(1)},\dots,Y^{(m)}\}$;.

\begin{enumerate}
	\item Compute the $z$-score of each item:
	      \[
		      \tilde Y^{(j)}_i
		      =\frac{Y^{(j)}_i-\mu_j}{\sigma_j},
		      \quad j=1,\dots,m,\; i=1,\dots,n .
	      \]

	\item For every non-empty subset \(S\subseteq\{1,\dots,m\}\), take the row-wise mean:
	      \[
		      Y^{\mathrm{comp}}_{i,S}
		      =\frac{1}{|S|}\sum_{j\in S}\tilde Y^{(j)}_i .
	      \]
\end{enumerate}

\noindent Each composite $Y^{\mathrm{comp}}_{S}$ then enters the usual model:
\[
	Y^{\mathrm{comp}}_{S}
	=F\bigl(X,\;Z_{\pi}\bigr)+\varepsilon
\]

\noindent analogous to Eq. \ref{objective_func}, with $X$ fixed and $Z_{\pi}\subseteq Z$ chosen by independant variable specification $\pi$. In this particular example, we highlight a specification containing two covariates and the $y1$ and $y2$ composites: \texttt{[`y1', `y2', `z1', `z2']}, and a specification containing only one covariate ($z4$) and a single operationalisation, non-composite of the dependent variable in the form of $y4$: \texttt{[`y4', `z4']}. If researchers are interested in running multiple fit commands -- one for each dependent variable without combinations of them -- \texttt{RobustiPy} provides \texttt{concat\_results} (import via \texttt{from robustipy.utils import concat\_results}) to stack \texttt{OLSResult} objects. For multiple binary dependents, run one \texttt{LRobust} fit per outcome and then stack the resulting objects with \texttt{concat\_results}.

\section{Results}\label{emp_and_sim}

All of the results in the following Sections \ref{sec:results_vanilla}-\ref{sec:results_y} --unless where specified -- are reported with 1,000 bootstrapped draws, ten folds, a seed of \texttt{192735}, and options of the HQIC information criteria for model selection, and the pseudo-R$^2$ metric for out-of-sample evaluation.\footnote{All of these options can be varied, along with further, additional ones; see Section \ref{sec:online_methods} and the accompanying \texttt{RobustiPy} documentation for further details.}$^,$\footnote{If draws, folds, seed, out-of-sample metric, or \texttt{n\_cpu} are omitted, \texttt{RobustiPy} prompts for them in interactive sessions; in non-interactive runs it uses defaults.}$^,$\footnote{When there are 128 or fewer specifications, subfigure f uses a dot plot; above this threshold, it uses a binned heatmap.} We provide the results of ten empirical examples which enunciate the five types of analysis as detailed in Sections \ref{sec_types:vanilla}-\ref{sec1:multiple_y}; some are in more detail than others in the main text, each with a corresponding notebook as part of our Online Supplementary Materials (i.e., \citealp{valdenegro_ibarra_2025_15700698}). All results are discussed in Section \ref{discussion} (and in particular, Table \ref{tab:results}).

\subsection{Vanilla computation}\label{sec:results_vanilla}

\begin{figure}[!t]
	\centering
	\includegraphics[width=\textwidth]{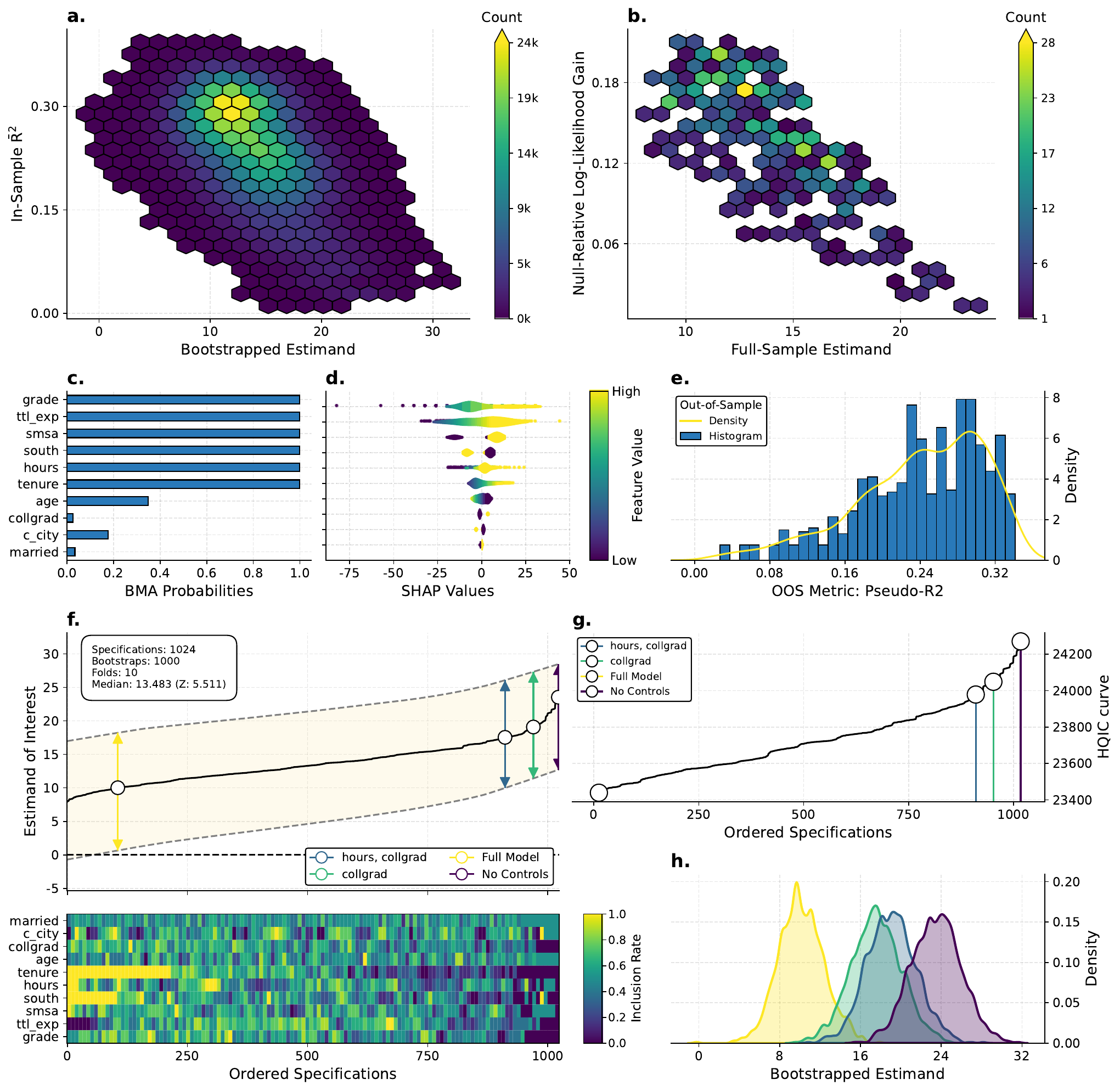}
	\caption{\textbf{Example output of \texttt{RobustiPy} utilising the \texttt{union.dta} dataset}. \small{Subfigure `\textbf{a.}' represents in-sample $\bar{R}^2$ fit against bootstrapped estimands, and `\textbf{b.}' represents the full-sample, non-bootstrapped estimands plotted against the null-relative log-likelihood gain (per observation) of the corresponding OLS models. Subfigure `\textbf{c.}' contains Bayesian Model Averages, and shares y-axis tick labels with `\textbf{d.}', which represents SHAP values. Subfigure `\textbf{e.}' shows the distribution of the chosen out-of-sample metric. In Subfigure `\textbf{f.}', median estimates across specifications are shown in solid black, while bootstrapped confidence intervals (with a default interval of the entire range) are shown in dashed grey (with optional LOESS). The colouring of specific lines, distributions, and markers in figures `\textbf{f.}', `\textbf{g.}', and `\textbf{h.}' represent the position of highlighted specifications of interest; \texttt{RobustiPy} will always show the position of the model with all variables (`Full Model') and the model with no control variables beyond the variable of interest (`No Controls'). The colormap can also be user-specified. Details of all metrics, methods, and features which create analysis such as this can be found in Section \ref{sec:online_methods}}.}
	\label{fig1}
\end{figure}

Our first example utilises the infamous \texttt{union.dta} dataset; a subsample of the National Longitudinal Survey of Young Women \citep{BLS1988}. %Women were surveyed in each of the 21 years 1968–1988, except for the six years 1974, 1976, 1979, 1981, 1984, and 1986). This canonical dataset is also well examined as an exemplary dataset in other work which considers model uncertainty and provides multiversal tools \citep{young2017model}. It is also a teaching resource -- used in undergraduate and graduate econometrics classes around the world -- to illustrate the estimation of treatment effects in observational data. Textbooks which utilise it include \cite{cameron2010microeconometrics} and \cite{acock2008gentle}, as just two examples. Through OLS regression, students and researchers estimate $\hat{\beta}$, which is interpreted as the average percentage difference in wages attributable to union membership, holding other factors constant.
This dataset is anonymised and publicly available. It contains information on things like wage (which, when logarithmically transformed, represents the dependent variable, often written as `$y_i$'), union membership (and the associated estimand of interest, `$\hat{\beta}_1$' in the canonical reduced form linear regression), and a multitude of often arbitrarily specified controls. Results are visualised in Figure \ref{fig1}; a simple OLS regression on the full sample yields an estimand of interest of 10.18; lower than `conventional estimates', which are nearer to a 15\% premium (Hirsch 2004). \cite{young2017model} estimates the (unweighted) median effect across the multiverse as 14.00; the median of our unweighted estimates across all samples with no bootstraps is 13.5, across all samples and bootstraps as 13.5, with a BIC-weighted estimate of 9.8 (as per the recommendation of \citealp{slez2019difference}). We are agnostic as to which number is most statistically valid, simply providing the tools by which a researcher can consider the range of plausible results.

We also undertake a `vanilla' application to criminology: a revisitation of \cite{ehrlich1973participation} based on a dataset from \cite{vandaele1987participation}, where the dependent variable is state-level crime rate, and the estimand of interest is income inequality (defined as the percentage of families earning below one-half of the median income). The results from the \texttt{.summary()} method -- inherent to \texttt{RobustiPy} -- are shown in \cbref{cb:criminality} and Supplementary Information Figure \ref{fig:S1}. This example is particularly interesting because -- as shown in other Supplementary Information Figures -- the direction of the estimand of interest actually changes across specification choice (e.g., from a minimum of -0.8743 to a maximum of 2.0337). The Supplementary Information (Figure \ref{fig:S2}) also contains a re-analysis of the `Colonial Origins of Comparative Development', based on the influential paper of \cite{acemoglu2001colonial}.

\begin{figure}[!t]
	\begin{codeblock}{Summarizing a re-analysis of criminality}{cb:criminality}
		\begin{minted}[baselinestretch=1]{python}
==============================
1. Model Summary
==============================
Model: OLS Robust
Inference Tests: Yes
Dependent variable: R
Independent variable: Inequality
Number of possible controls: 7
Number of draws: 1000
Number of folds: 10
Number of specifications: 128
==============================
2.Model Robustness Metrics
==============================
2.1 Inference Metrics
==============================
Median β (specs, no resampling): 0.6976 (null-calibrated p: 0)
Median β (bootstraps x specs): 0.6134
Min β (specs, no resampling): -0.2304
Min β (bootstraps x specs): -0.8743
Max β (specs, no resampling): 1.0851
Max β (bootstraps x specs): 2.0337
AIC-weighted β (specs, no resampling): 0.7754
BIC-weighted β (specs, no resampling): 0.7296
HQIC-weighted β (specs, no resampling): 0.7588
Share significant (specs, no resampling) [descriptive]: 0.7500
Share significant (bootstraps x specs) [descriptive]: 0.7293  (p: 0.001998)
Share β>0 (specs, no resampling) [descriptive]: 0.8750
Share β>0 (bootstraps x specs) [descriptive]: 0.8632 (p: 0.2498)
Share β<0 (specs, no resampling) [descriptive]: 0.1250
Share β<0 (bootstraps x specs) [descriptive]: 0.1368  (p: 0.2498)
Share β>0 & significant (specs, no resampling) [descriptive]: 0.7500
Share β>0 & significant (bootstraps x specs) [descriptive]: 0.7141 (p: 0.001998)
Share β<0 & significant (specs, no resampling) [descriptive]: 0.0000
Share β<0 & significant (bootstraps x specs) [descriptive]: 0.0152 (p: 0.001998)
Stouffers Z = 3.6834  (two-sided p = 0.000999)
==============================
2.2 In-Sample Metrics (Full Sample)
==============================
Min AIC: 427.4263, Specs: ['Expenditure', 'Wealth', 'Ed', 'Age']
Min BIC: 438.5272, Specs: ['Expenditure', 'Wealth', 'Ed', 'Age']
Min HQIC: 431.6036, Specs: ['Expenditure', 'Wealth', 'Ed', 'Age']
Max Relative Log-Likelihood Gain (per obs): 0.664, Specs: ['Expenditure', 'Males', 'N',
                                                           'Age', 'Unemployment', 'Ed',
                                                           'Wealth']
Min Relative Log-Likelihood Gain (per obs): 0.0163, Specs: []
Max Raw Log Likelihood: -206.7716, Specs: ['Expenditure', 'Males', 'N', 'Age',
                                           'Unemployment', 'Ed', 'Wealth']
Min Raw Log Likelihood: -237.2149, Specs: []
Max Adj-R2: 0.6922, Specs: ['Expenditure', 'Males', 'Age', 'Ed', 'Wealth']
Min Adj-R2: -0.0310, Specs: ['Unemployment', 'Age']
==============================
2.3 Out-Of-Sample Metrics (pseudo-r2 averaged across folds)
==============================
Max Average: 0.4925, Specs: ['Expenditure', 'Unemployment', 'Ed', 'Age']
Min Average: -0.5057, Specs: ['Males', 'Unemployment', 'Ed', 'Age']
Mean Average: 0.12
Median Average: 0.2026
\end{minted}
	\end{codeblock}
\end{figure}

\begin{figure}[!t]
	\centering
	\includegraphics[width=\textwidth]{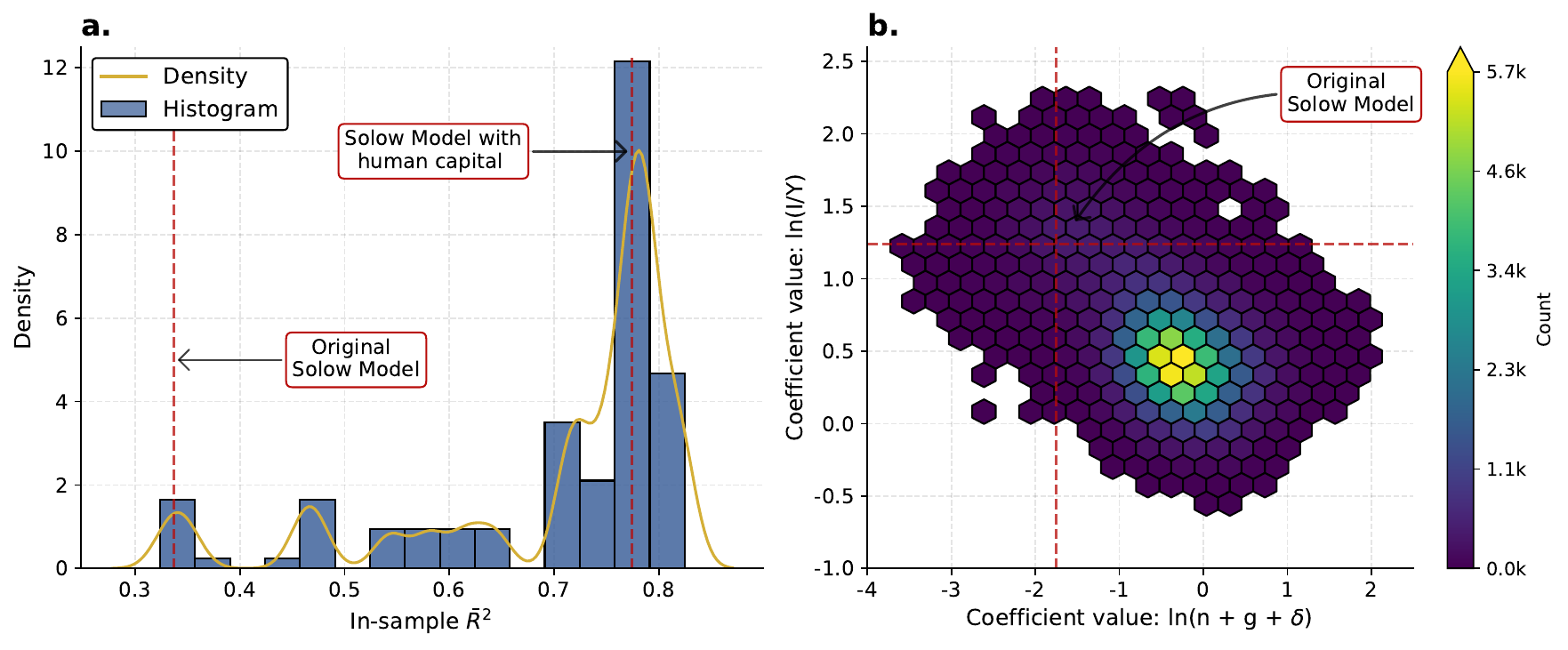}
	\caption{\textbf{Replicating and extending \cite{mankiw1992contribution} with \texttt{RobustiPy}}. With regards to subfigure `\textbf{a.}', the distribution of $\bar{R}^2$ is identical for each of the two orderings of $\overleftrightarrow{\mathbf{X}}$, as the model space is otherwise identical across the two results objects. The right subfigure (`\textbf{b.}') is a composite of two runs of \texttt{RobustiPy}, with the two rotated $\overleftrightarrow{\mathbf{X}}$ arrays. The figure illustrates the substantive finding of \cite{mankiw1992contribution}: the basic Solow model without human capital exhibits systematically lower explanatory power, while augmenting the model with a human capital proxy yields markedly higher $\bar{R}^2$ values. The hexbin representation in panel `\textbf{b.}' makes clear that the coefficients vary enormously in comparison to the small number of values reported by \cite{mankiw1992contribution}.}
	\label{fig:mrw}
\end{figure}

\subsection{Array of never changing variables}\label{sec:results_constants}

We next detail the case where we have an array of predictor names that we always want to include in the model space, but never vary. In terms of \texttt{RobustiPy} functionality, this simply involves passing an array of predictor names ($\overleftrightarrow{\mathbf{X}}$) which has length greater than one, with the array of variable controls specified as before ($\overleftrightarrow{\mathbf{Z}}$). This involves using updated variables from the Penn World Table (version 10.01, see \citealp{feenstra2015next}) to revisit \cite{mankiw1992contribution} which contributes to the `Empirics of Economic Growth'. We map modern data from 73 of the 98 countries in the original 1988 sample, using observed values of all parts of the $\log(n+g+\delta)$ and $\log(\frac{I}{Y})$ which are canonical to the Solow Growth Model. We also include various possible extensions -- as done in \cite{mankiw1992contribution} -- such as including logged metrics of human capital, and previous values of GDP in levels. By running \texttt{RobustiPy} twice -- with $\overleftrightarrow{\mathbf{X}} = [\log(n+g+\delta), \log(\frac{I}{Y})]$ and $\overleftrightarrow{\mathbf{X}} = [\log(\frac{I}{Y}), \log(n+g+\delta)]$, both of which we always want in our core model -- we are able to recover two large sets of bootstrapped estimates of the relevant estimands accompanying the first variable in each of these two arrays. This allows us to extract estimates from the \texttt{results} object of \texttt{RobustiPy} to assess whether the signs are of equal size and opposite magnitude. \texttt{RobustiPy} also allows us to visualise the distribution of the $\bar{R}^2$ values (a key test of the Solow Growth Model); both are shown in Figure \ref{fig:mrw}, with full \texttt{RobustiPy} outputs for the two runs shown in Figures \ref{fig:S3}-\ref{fig:S4}.

\begin{figure}[!t]
	\centering
	\includegraphics[width=\textwidth]{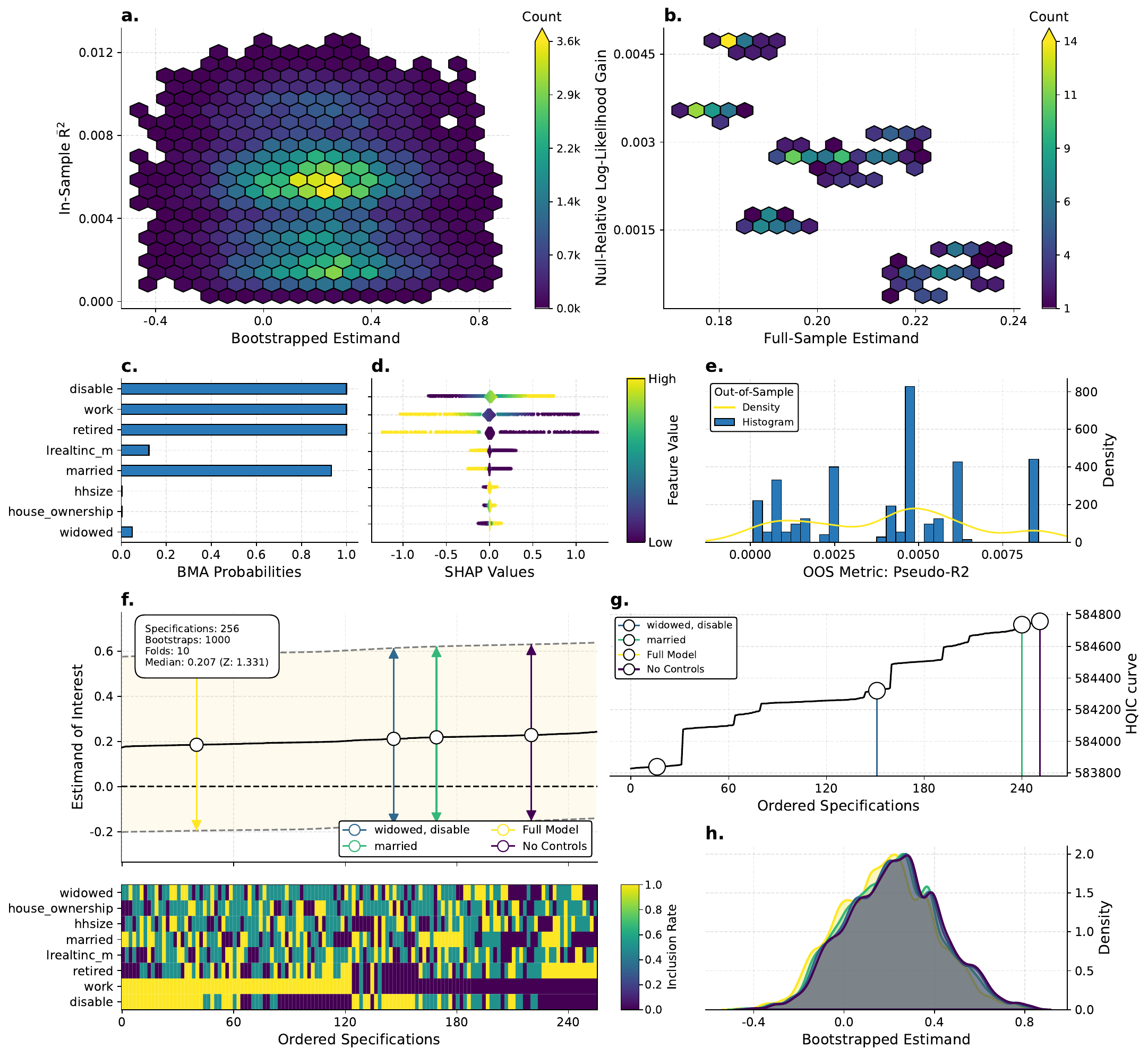}
	\caption{\textbf{Example output of \texttt{RobustiPy} with both a grouping variable and an array of never changing variables}. The dependent variable is a measure of general health (GHQ12), and the estimand of interest ($\hat{\beta}$) is associated with spending on adult social care.}
	\label{figure:asc}
\end{figure}

\subsection{Fixed effect OLS}\label{results:fixed_effects}

Regarding fixed-effect estimation, our empirical replication comes from \cite{zhang2021longitudinal}, which takes a longitudinal approach to local government spending on adult social care and carers’ subjective well-being in England. Specifically, we analyse the effect of a Local Authority measure of Adult Social Care Spending on a General Health Questionnaire (GHQ12) metric from the United Kingdom Household Longitudinal Survey (see Appendix I, column (1) of \citealp{zhang2021longitudinal}). A static group of predictors (i.e., inside $\overleftrightarrow{\mathbf{X}}$) includes seven sociodemographic and health variables, alongside thirteen year fixed effect dummy variables, whereas our array of controls -- which do vary -- contains a further eight variables. Importantly, we document the inclusion of a grouping variable (`pipd'), common in longitudinal survey analysis. Figure \ref{figure:asc} shows the results, with the median value of the estimand of interest -- 0.2069 for all specifications and all bootstrapped resamples, 0.2016 for all specifications and no resampling, and 0.1825, 0.1818 and 0.1826 for AIC, BIC, and HQIC weighted estimates -- obtained from an unreported print-out analogous to \cbref{cb:criminality}. This result is extremely close to that reported in the original paper (0.1825). Given that the size of this feature space is so large, had we not specified a larger $\overleftrightarrow{\mathbf{X}}$ (an array of always present predictors), \texttt{RobustiPy} would have had to estimate a relatively intractable control-inclusion space of $2^{d_Z}=268{,}435{,}456$ specifications, where $d_Z=28$ is the number of candidate control variables. We could have alternatively used our method of resampling $\Pi_{\overleftrightarrow{\mathbf{Z}}}$. Results are again also shown in Table \ref{tab:results}. A complementary pedagogical example -- with the Understanding Society Longitudinal Teaching Dataset (Waves 1-9, from October 2021, see \citealp{understanding_society_2024}) is provided in Supplementary Figure \ref{fig:S5}, where the estimand of interest is `\texttt{sf1\_dv}' (general health) and a fixed effect grouping is based on a unique cross-wave person identifier.

\subsection{Binary dependents}\label{sec:results_binary}

We next describe an example of how \texttt{RobustiPy} can be used when our dependent variable is binarized as part of an application common to the medical sciences. In particular, we utilise the Framingham Heart Survey \citep{dawber1951epidemiological} for the estimation of models related to a dependent variable termed `TenYearCHD': a binary indicator equal to one if a study participant experienced a coronary heart disease (CHD) event within ten years of their baseline examination, and zero otherwise. This is used in multiple high-profile papers such as \cite{wilson1998prediction}, and as part of a composite in papers such as \cite{d2008general}. The dataset includes over 4,240 records across sixteen features. We fit our \texttt{LRobust} class with `TenYearCHD' as our target dependent, `BMI' as the single element of $\overleftrightarrow{\mathbf{X}}$ for which we calculate estimands of interest, and we use `age', `male', `totChol', `sysBP', `BPMeds', `currentSmoker', `cigsPerDay', `diabetes', `heartRate', and `glucose' as elements of our array of controls ($\overleftrightarrow{\mathbf{Z}}$). This example accepts the `\texttt{oddsratio}' boolean into the \texttt{.plot()} command on the results object to calculate odds ratios, with results shown in Figure \ref{figure:framingham}. Our online replication archive also includes models fit to a Kaggle dataset on a binarized measure of airline customer satisfaction (from an undisclosed airline); the resulting coefficient estimates indicate how \texttt{RobustiPy} can be used as a method for model selection and averaging, and are presented in Figure \ref{fig:S_airline} and, again, summarized in Table \ref{tab:results} (which also contains \texttt{RobustiPy} results from estimating the problem as a Linear Probability Model, i.e., with the \texttt{OLSRobust} class).

\begin{figure}[!t]
	\centering
	\includegraphics[width=\textwidth]{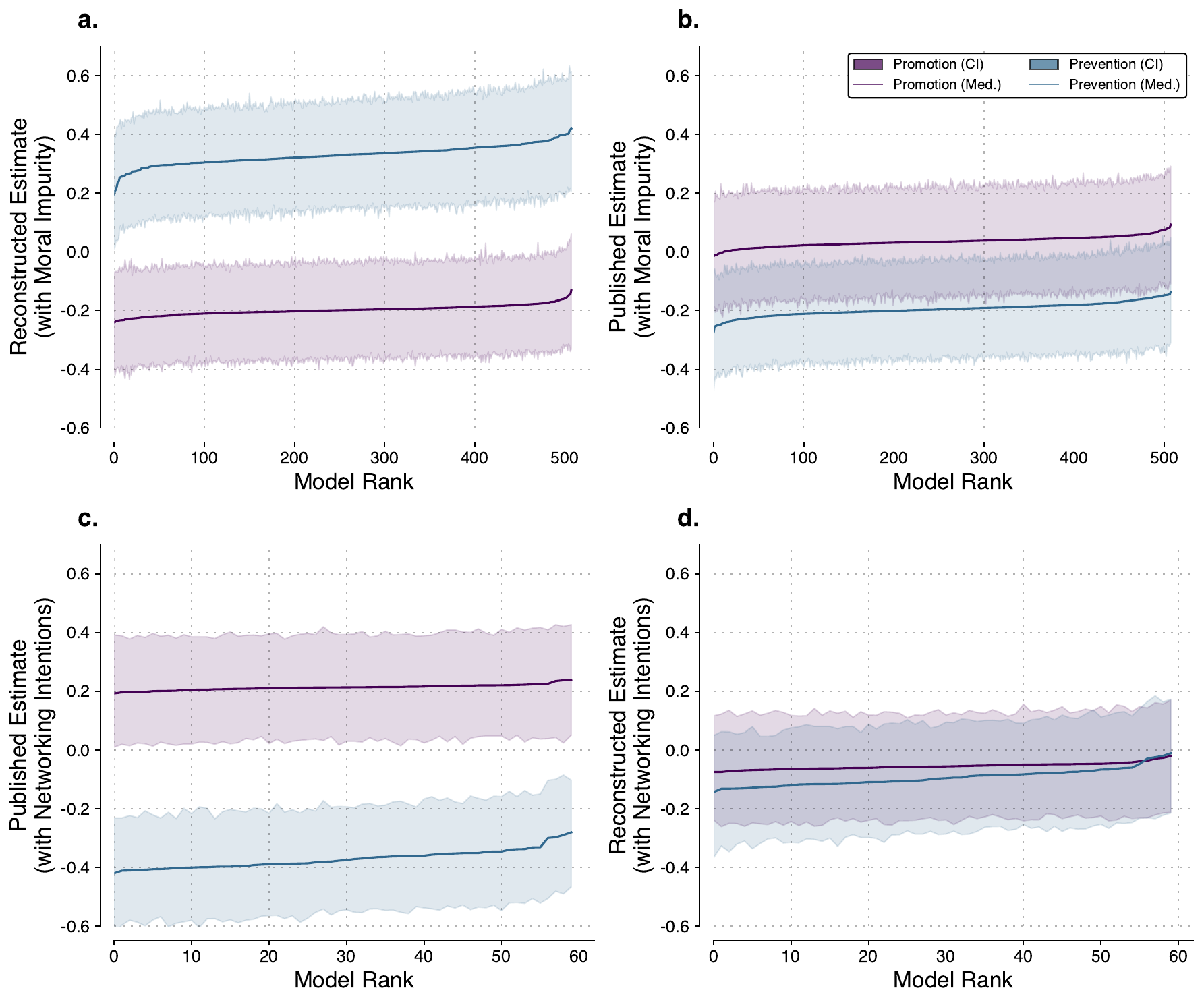}
	\caption{\textbf{\small{Output of \texttt{OLSRobust} using multiple dependent variables.}} In this example we use a custom plotting function which accepts input from the output results object of the \texttt{OLSResult} class -- comparing multiple competing versions of $\Pi$ -- based on \cite{gino2020connect}. The ``Promotion'' and ``Prevention'' condition refer to the experimental conditions of the original study.}
	\label{gino_fig}
\end{figure}

\subsection{Multiple dependents}\label{sec:results_y}

\begin{figure}[!t]
	\centering
	\includegraphics[width=\textwidth]{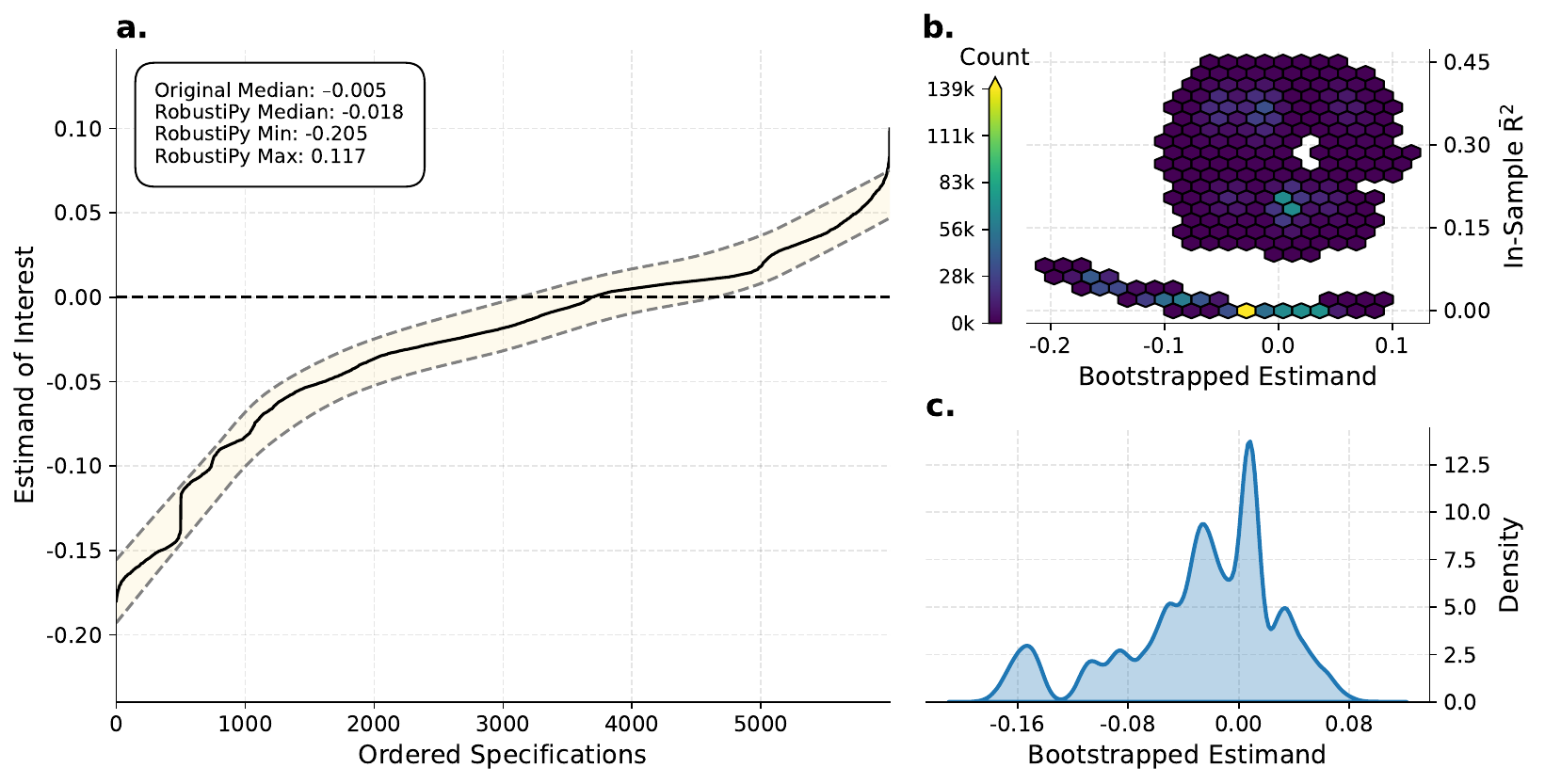}
	\caption{\textbf{Example output of \texttt{RobustiPy} utilising multiple dependent variables}.
		This example is taken from \cite{orben2019association}. It involves multiple dependent variables, and rescaled independent variables. It also involves a range of different predictor sets, and multiple runs of \texttt{RobustiPy} stacked together using `\texttt{.concat\_results}'. It largely substantiates the results of the original study, showing that the estimand of interest straddles both above and below zero. This example is run with 250 combinations of multiple dependents and 250 bootstrap resamples due to computational cost, showing the flexibility of \texttt{RobustiPy} to adapt to computational power.}
	\label{fig:Orben1}
\end{figure}

We next present the case whereby a study might have multiple candidates for a dependent variable that measures the same underlying concept. While this idea might be uncommon in some disciplines, having a composite target variable that combines multiple measures is extremely common practice in myriad academic disciplines. The most common case of multiple target variables arises from the use of multi-item measures in surveys. Here, we attempt to replicate the results of Study 3a in \cite{gino2020connect}, alongside further replication analysis conducted by DataColada for the same study. In their analysis (see \citealp{simmons2024harvard_gino_report}), DataColada alluded to the manipulation of the results through the alteration of the data used in the originally published data and analyses, ultimately leading to a retraction of the original study (see \citealp{harvard_gino_report_2024}). Here, we attempt to replicate the effect of the experimental condition on measures of moral impurity (seven items) and networking intentions (four items), separating the datasets used in both the originally published study, and that which has been reconstructed by \cite{harvard_gino_report_2024}. In the case of multiple dependent measures, \texttt{RobustiPy} combines them into all possible composite combinations (i.e. average of the standardized z scores of the measures), including single-item combinations. To ensure comparability between items, \texttt{RobustiPy} uses $z$-scores rather than raw measures (see Section \ref{sec1:multiple_y}). Finally, to replicate an ANOVA/ANCOVA analysis, we convert the categorical column containing the experimental condition group membership into dummy variables and use them as predictors in each case. The results are shown in Figure \ref{gino_fig}. The resultant outputs from \texttt{RobustiPy} are used to plot specification curves based on the published dataset which would appear to support the main hypothesis of the study, as it shows that any combination of the outcome variable `moral impurity' has a significant effect from the experimental condition. However, results using the reconstructed data show that the effect is much weaker and in the opposite direction than hypothesised. This example illustrates the flexibility of \texttt{RobustiPy}, demonstrating how it can be used to replicate both ANOVA and ANCOVA analyses -- even though these are not explicitly implemented in the library -- while also enabling the integration of custom visualisations to present results. It also emphasizes the capacity of \texttt{RobustiPy} to act as a tool for replication. More broadly, this example shows how \texttt{RobustiPy} can be employed as a tool for auditing existing studies by systematically exploring the robustness of published findings. By taking the output of the `\texttt{.get\_results()}' object, \texttt{RobustiPy} enables a multitude of custom comparisons beyond that which a single run permits. We also use \texttt{RobustiPy} to replicate \cite{orben2019association} which aims to understand the `association between adolescent well-being and digital technology use'. In this article, data from well-known adolescent well-being panel studies is used across a series of measures of well-being (dependent variables) and digital use (independent variables). Under this setting, the authors opt for a specification curve analysis approach to obtain a distribution of the effect of many forms of technology use across many measures of well-being; we replicate one of the main analyses, where the results -- which are also a function of \texttt{RobustiPy}'s ability to rescale $\overleftrightarrow{\mathbf{X}}$ and $\overleftrightarrow{\mathbf{Z}}$ --  are seen in Figure \ref{fig:Orben1} and Table \ref{tab:results}.

\section{Discussion}\label{discussion}

Table \ref{tab:results} outputs all results from our ten empirical replications. It explicates the enormous variation in both the estimand of interest -- across a researcher's set of reasonable choices -- and also the variation in out-of-sample accuracy across specification spaces. It also shows, for example, how we can consider whether coefficient values are approximately equal and opposite across various different realisations of an estimand (weighted or unweighted), as done by \cite{mankiw1992contribution} in their test of the Solow Model. It emphasizes the ability of \texttt{RobustiPy} to act as a tool for the evaluation of research already published in the scientific record, not least through its ability to highlight specific specifications of interest in the visualisation; it acts as an essential audit tool when the validity of research is under question (such as through the retraction of \citealp{gino2020connect}). While not emphasized elsewhere in this manuscript, \texttt{RobustiPy} also outputs exact specifications which maximize out-of-sample evaluation metrics, and which specifications minimize certain information criteria; this function is immensely useful for helping to maximize model fit (see \cbref{cb:criminality} for an example of this). While Table \ref{tab:results} is compiled from the \texttt{.summary()} method, this is just one of the three ways by which \texttt{RobustiPy} provides output; it also provides a large number of (separable) visualisations alongside \emph{all} raw estimates (which can be used for bespoke downstream analyses).

\afterpage{%
	\begin{landscape}
		\begin{table}[!p]
			\centering
			% make a box exactly \textwidth wide
			\begin{minipage}{1.4\textwidth}
				% resize tabular to fill the full width of that minipage
				\resizebox{\textwidth}{!}{%
					\begin{tabular}{l ccc | cccc | cc | cc | c}
						\toprule
						                                                    & \multicolumn{3}{c|}{\textbf{All specs, with resamples}}
						                                                    & \multicolumn{4}{c|}{\textbf{All specs, no resamples}}
						                                                    & \multicolumn{2}{c|}{\textbf{Stouffer's}}
						                                                    & \multicolumn{2}{c|}{\textbf{OOS R${^2\ddagger}$}}
						                                                    &                                                                                                                                                                                                                                                                                      \\
						\cmidrule(lr){2-5} \cmidrule(lr){6-9} \cmidrule(lr){10-11} \cmidrule(lr){12-13}
						\textbf{Replication}
						                                                    & Min                                                     & Median & Max    & Min                      & Median                                                                    & Max   & BIC$^\dagger$ & Z       & p     & Min   & Max
						                                                    & \textbf{Notes}                                                                                                                                                                                                                                                                       \\ \midrule

						\color{darkblue} Union Dataset\color{black}
						                                                    & -1.2                                                    & 13.5   & 31.7   & 8.10                     & 13.5                                                                      & 23.7  & 9.8           & 5.5     & 0.001 & 0.0   & 0.3  & See \cite{young2017model}                \\

						\cite{acemoglu2001colonial}                         & 0.23                                                    & 0.44   & 0.62   & 0.39                     & 0.44                                                                      & 0.53  & 0.42
						                                                    & 25.4                                                    & 0.001  & 0.59   & 0.7                      & Constant in $ \overleftrightarrow{\mathbf{X}}$                                                                                                                                \\

						\cite{acemoglu2001colonial}                         & 0.23                                                    & 0.44   & 0.62   & 0.39                     & 0.81                                                                      & 1.23  & 0.42
						                                                    & 25.4                                                    & 0.001  & -0.43  & 0.7                      & Constant in $ \overleftrightarrow{\mathbf{Z}}$                                                                                                                                \\

						\cite{vandaele1987participation}                    & -0.87                                                   & 0.61   & 2.03   & -0.23                    & 0.70                                                                      & 1.09  & 0.73          & 3.68    & 0.001 & -0.51 & 0.49 & See also \cite{ehrlich1973participation} \\

						\cite{mankiw1992contribution}                       & -0.55                                                   & 0.46   & 2.24   & 0.26                     & 0.41                                                                      & 1.31  & 0.30
						                                                    & 2.90                                                    & 0.021  & 0.31   & 0.8                      & $\overleftrightarrow{\mathbf{X}}=[\log\!\tfrac{I}{Y},\,\log(n+g+\delta)]$                                                                                                     \\

						\cite{mankiw1992contribution}                       & -3.59                                                   & -0.34  & 1.99   & -1.82                    & -0.21                                                                     & 0.27  & -0.17
						                                                    & -0.98                                                   & 0.350  & 0.31   & 0.8                      & $\overleftrightarrow{\mathbf{X}}=[\log(n+g+\delta),\,\log\!\tfrac{I}{Y}]$                                                                                                     \\

						\cite{understanding_society_2024}                   & 0.01                                                    & 0.01   & 0.01   & 0.01                     & 0.01                                                                      & 0.01  & 0.01          & 37.1    & 0.001 & 0.01  & 0.01 & Grouping variable                        \\

						\cite{zhang2021longitudinal}                        & -0.46                                                   & 0.21   & 0.85   & 0.17                     & 0.20                                                                      & 0.24  & 0.18
						                                                    & 1.33                                                    & 0.330  & 0.0001 & 0.009                    & Multiple $\overleftrightarrow{\mathbf{X}}$                                                                                                                                    \\

						%IN CODE THE STOUFFER IS NAN BECAUSE NOT IMPLEMENTED IN LR?!!?
						\cite{dawber1951epidemiological}                    & -0.06                                                   & 0.03   & 0.10   & 0.0                      & 0.02                                                                      & 0.06  & 0.01
						                                                    & n/a                                                     & n/a    & 0.01   & 0.1                      & \texttt{LRobust}                                                                                                                                                              \\

						\cite{dawber1951epidemiological}                    & -0.01                                                   & 0.00   & 0.01   & -0.0                     & 0.00                                                                      & 0.01  & 0.00
						                                                    & 2.06                                                    & 0.064  & 0.064  & 0.09                     & \texttt{OLSRobust}                                                                                                                                                            \\

						\color{darkblue}Passenger Satisfaction\color{black} & 0.01                                                    & 0.01   & 0.02   & 0.01                     & 0.01                                                                      & 0.02  & 0.01          & n/a     & n/a   & 0.02  & 0.23 & \texttt{LRobust}                         \\

						\color{darkblue}Passenger Satisfaction\color{black} & 0.00                                                    & 0.00   & 0.00   & 0.00                     & 0.00                                                                      & 0.00  & 0.00          & 38.2    & 0.001 & 0.44  & 0.49 & \texttt{OLSRobust}                       \\

						%1
						\cite{gino2020connect}                              & -0.61                                                   & -0.20  & 0.22   & -0.24                    & -0.2                                                                      & -0.14 & n/a           & -49.8   & 0.001 & 0.02  & 0.07 & Promotion, published                     \\

						%2
						\cite{gino2020connect}                              & -0.19                                                   & 0.33   & 0.85   & 0.20                     & 0.33                                                                      & 0.42  & n/a           & 79.6    & 0.001 & 0.86  & 0.99 & Prevention, published                    \\

						%3
						\cite{gino2020connect}                              & -0.41                                                   & 0.03   & 0.49   & -0.01                    & 0.03                                                                      & 0.09  & n/a           & 8.60    & 0.010 & -0.0  & 0.02 & Promotion, original                      \\

						%4
						\cite{gino2020connect}                              & -0.63                                                   & -0.2   & 0.23   & -0.27                    & -0.2                                                                      & -0.14 & n/a           & -45.9   & 0.001 & -0.0  & 0.02 & Prevention, original                     \\

						%5
						\cite{gino2020connect}                              & -0.21                                                   & 0.21   & 0.64   & 0.20                     & 0.21                                                                      & 0.24  & n/a           & 17.9    & 0.001 & 0.04  & 0.06 & Promotion, published$^*$                 \\

						%6
						\cite{gino2020connect}                              & -0.77                                                   & -0.37  & 0.11   & -0.41                    & -0.37                                                                     & -0.28 & n/a
						                                                    & -29.4                                                   & 0.001  & 0.04   & 0.06                     & Prevention, published$^*$                                                                                                                                                     \\

						%7
						\cite{gino2020connect}                              & -0.44                                                   & -0.05  & 0.38   & -0.07                    & -0.06                                                                     & -0.03 & n/a           & -4.45   & 0.001 & -0.01 & 0.01 & Promotion, original$^*$                  \\

						%8
						\cite{gino2020connect}                              & -0.54                                                   & -0.09  & 0.37   & -0.14                    & -0.09                                                                     & -0.01 & n/a           & -7.26
						                                                    & 0.001                                                   & -0.01  & 0.01   & Prevention, original$^*$                                                                                                                                                                                 \\

						\cite{orben2019association}                         & -0.20                                                   & -0.02  & 0.12   & -0.18                    & -0.02                                                                     & 0.1   & n/a           & -531.29 & 0.004 & n/a   & n/a  & Multiple dependents                      \\
						\bottomrule
					\end{tabular}%
				}
				% now caption is exactly \textwidth wide, flush left with the table
				\caption{\textbf{Comparative analysis of different datasets with various specifications and results.} All data is publicly available, and all results are obtained using \texttt{RobustiPy}. $^\dagger$ refers to estimands weighted by the Bayesian Information Criterion (\texttt{RobustiPy} also calculates the AIC and HQIC, with ready extension to other Information Criteria). $^\ddagger$ represents the out-of-sample R$^2$ metric (\texttt{RobustiPy} offers various alternate out-of-sample model evaluation metrics; see Section \ref{methods:oos}). $^*$ indicates the use of networking intention items. Please see our Data Availability Statement for information on accessing each of these datasets. Information criterion weighted values are not appropriate for studies with multiple dependent variables;  the log likelihood is defined with respect to the probability density (or mass) of the specific dependent variable, and multiple different dependent variables induce different sample spaces and likelihood functions, rendering their log-likelihoods incommensurable. It is for this reason that our visualisations for this type of analysis omit panels that rely on likelihood-comparable metrics. The results for \cite{orben2019association} are run with 250 resampled draws across $\overleftrightarrow{\mathbf{Y}}$, and with 250 bootstrapped draws.}\label{tab:results}
			\end{minipage}
		\end{table}
	\end{landscape}
}

\begin{figure}[!t]
	\centering
	\includegraphics[width=\textwidth]{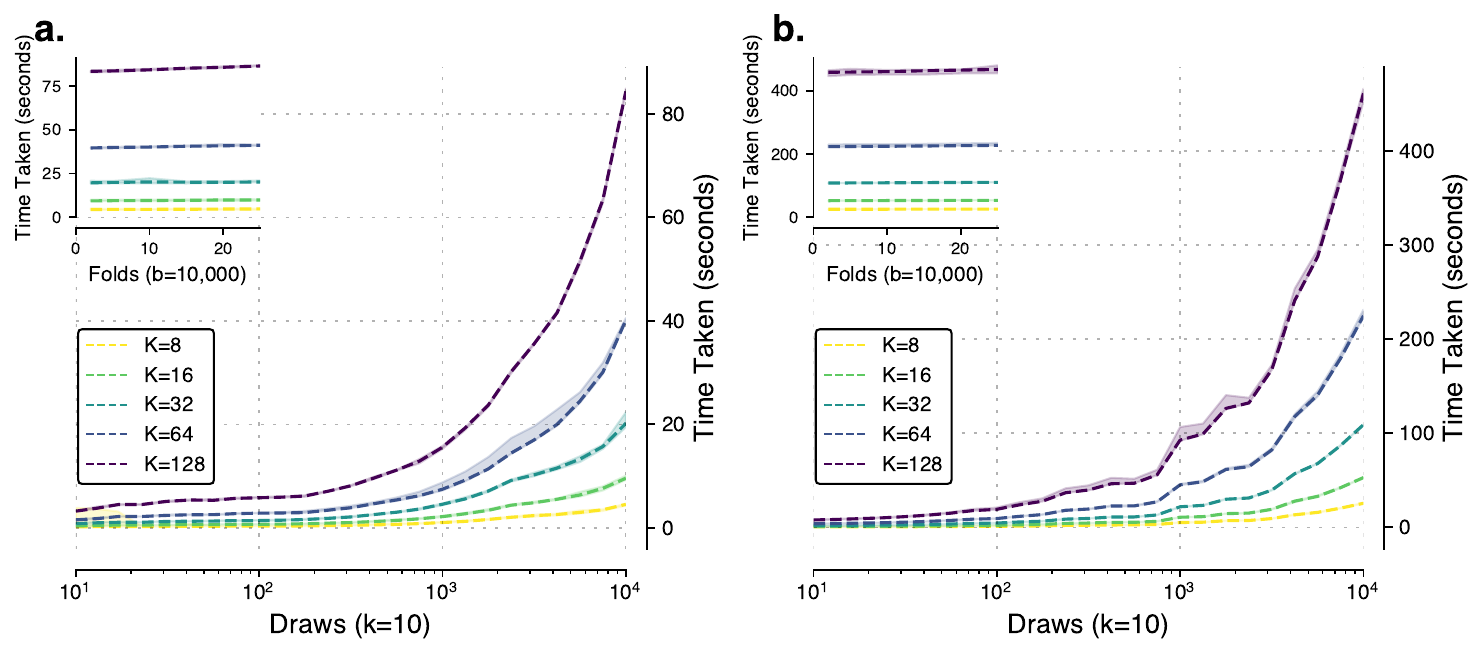}
	\caption{\textbf{Time profiling of \texttt{RobustiPy} with simulated data}. Subfigure `\textbf{a.}' represents the use of the \texttt{OLSRobust} class, and `\textbf{b.}' the case of the \texttt{LRobust} class. Shaded areas represent the range across ten trials (i.e., filled between minimum and maximum runtime for that exact parameterisation of the simulation). Dashed lines represent the median. Insets show variation across number of folds ($k$), while main subfigures represent variation across number of draws ($b$).}
	\label{fig:time_profiler}
\end{figure}

We also time profile \texttt{RobustiPy}. It involves two simple simulations of data generating processes (see Online Supplementary Material; \citealp{valdenegro_ibarra_2025_15700698}), and contains an example of each of both the \texttt{OLSRobust} and \texttt{LRobust} classes. We generate 25 integer values on a log scale between 10 and 10000 which represent a varying number of draws ($b$). We use the `\texttt{.fit()}' method with 2, 5, 10, 15, 20, and 25 folds ($k$). We use 3, $\hdots$, 7 controls (which results in groups of $\{2^3, \hdots, 2^7\}$ specifications). For each combination of folds, draws, and number of specifications, the problem is estimated ten times. The seed is unchanged for each iteration, and we use twenty cores of a 14900F processor, 96GB of DDR5 Random Access Memory with \texttt{Python} 3.12 installed on an Ubuntu 24.10 operating system. This results in approximately 671,512,080 regressions being estimated for our time profiling in total, including those done to create joint-inference tests (see Section \ref{sec:online_methods}). With the results shown in Figure \ref{fig:time_profiler}, we hypothesize that \texttt{RobustiPy} operates as an approximately $O(K(2b+k))$ process.

We aim to provide an accessible tool for researchers to run the most common associative models in an era when computational power is vast. We offer this easy-to-use interface in the hope that transparent results‐reporting will spread and become the default standard for statistical analysis. By lowering the user-experience barrier, we aim to accelerate the adoption—and routine practice—of robust modeling. We include as many tools, tests, and features as possible whilst attempting to maintain the simplest possible interface in order to encourage researchers to explore the robustness of the models which test their research hypotheses. Through the analysis of multiple aggregated runs of RobustiPy, researchers can also make assertions about whether to include transformed variables, such as $Z^2$ instead of $Z$.

Importantly, our application is highly modular and easily extendible.\footnote{We are constantly looking for suggestions with regards to new features to incorporate. Please contact the corresponding authors to discuss!} We envision tools such as \texttt{RobustiPy} as a necessity for a large majority of empirical workflows, yet there are some natural limitations which warrant further elaboration. First, while \texttt{RobustiPy} offers a comprehensive array of tools to conduct various forms of model uncertainty, many aspects of the process rely heavily on decisions made by the researcher. The selection of a defensible array of control variables, the choice of the underlying estimator, and ultimately the hypothesis to be tested are all left to the researcher's discretion; spurious or poorly specified models will still yield spurious and poorly specified results, even when implemented using tools like \texttt{RobustiPy} (despite the fact that we also weight by information criteria, as advocated by \citealp{slez2019difference} in response to \citealp{young2019difference}). Our model‑averaged estimates should be interpreted as conditional on the assumptions embedded in the candidate models and in the information‑criterion weighting scheme. AIC/BIC weights are derived from an assumed likelihood, large‑sample approximations, and a specific objective that balances relative fit against complexity; they are not designed to prioritize unbiasedness. Consequently, selecting or averaging “preferred” estimates can still transmit misspecification, and it does not address omitted variable bias, which is a separate threat arising from excluded covariates or incorrect functional form. For this reason, the weighted estimate is best viewed as a pragmatic summary of model uncertainty rather than a claim of bias‑free inference. Moreover, information criteria reward predictive adequacy under the assumed model class, not causal identification or robustness to unmeasured confounding, so a higher weight does not guarantee greater substantive validity. This implies that differences across specifications may reflect both genuine sensitivity and the limits of the model class itself, and that small shifts in covariate choice or functional form can meaningfully change weighted summaries. The central interpretive point is that model averaging can stabilize estimates when evidence is diffuse, but it cannot substitute for theory‑driven model construction or for designs that directly mitigate bias. These limits should temper over‑interpretation of precise point estimates and encourage emphasis on uncertainty and sensitivity. Our application does not presently conduct mis-specification tests, for example, which is a natural extension. In particular, formal diagnostics for functional form, heteroskedasticity, serial correlation, and omitted-variable bias remain external to the current implementation and must be undertaken separately by the researcher.

Weighting only has meaning within the space of models to which it is applied, so the central practical question is whether the candidate set spans a reasonable and defensible specification space. Any weighting rule will reflect—and potentially amplify—the structure built into that space, including decisions about covariates, functional forms, identification strategies, and preprocessing. In this sense, there is no universally superior weighting procedure: the most appropriate choice depends on the available inputs and the inferential goal, but all of them inherit the limits of the model space they operate over \citep{young2025multiverse}. Relatedly, fit‑based selection can privilege models that align poorly with theoretical expectations, underscoring the need to justify the specification space itself \citep{auspurg2025robustness}. However, ideological definitions drive what is constituted as a `reasonable' space, and that itself is contested \citep{ganslmeier2025reply}. The key implication is that careful construction and transparent reporting of a defensible model space is at least as important as the choice of weighting rule.

While analysis of this kind provides a computationally powerful framework for testing the robustness of statistical models, readers should not assume that it is the ultimate or only approach for such assessments. Like any method, it is  most useful when applied critically and in conjunction with substantive theoretical reasoning. The natural extensions to \texttt{RobustiPy} are limitless. % updates are the ability to create interaction terms among a subset of the control variables.
We plan to explore incorporating Laplacian approximations into \texttt{RobustiPy} by expanding each model’s log-posterior around its maximum a-posteriori estimate, using the resulting Gaussian integral as an efficient proxy for the marginal likelihood, thereby enabling scalable comparisons across large model spaces \citep{tierney1986accurate} in order to reduce compute costs \citep{strubell2019energy}. Other natural extensions involve the incorporation of additional estimators beyond the OLS, Logit, and Fixed Effects models currently supported. There is also pioneering work which attempts to equivocate estimators \citep{young2021functional}, much of which will be incorporated as we continue to support enhancements to the \texttt{RobustiPy} framework over time. %These include—but are not limited to—t-tests, ANOVA, MANOVA, ANCOVA, MANCOVA, and Probit models, all of which can be considered natural extensions of the existing model classes. Finally, a port to the R programming language is on the agenda, along with the development of a graphical user interface for a subset of the functionality.

\section{Resource Availability}\label{availability}

\textbf{Code Availability Statement:} The extensive code library which accompanies this work can be found at \href{https://www.github.com/RobustiPy}{github.com/RobustiPy}, on the Python Package Index which avails an install as simple as a \texttt{pip install robustipy} command, and is also indexed into Zenodo as `Online Supplementary Material'. This is referenced and cited in the text as \cite{valdenegro_ibarra_2025_15700698}. Documentation is available at \href{https://robustipy.readthedocs.io/}{robustipy.readthedocs.io}.\\

\noindent \textbf{Data Availability Statement:} All but three of our empirical examples utilise publicly available information, and those three are in general highly accessible upon requests to the relevant parties. In our replication files \citep{valdenegro_ibarra_2025_15700698}, two of these three require a simple -- an immediately granted -- application to the UK Data Service; these are \texttt{empirical\_type5\_orben.ipynb} and \texttt{empirical\_type3\_ukhls.ipynb}. One -- \texttt{empirical\_type3\_asc.ipynb} -- requires a request to the authors of \cite{zhang2021longitudinal}. The other seven are entirely publicly available, but we re-share no data as part of our library in order to be maximally respectful of the relevant data licenses (even if some permit re-sharing and are in the public domain). Of these, our replication files automatically scrape necessary data on the fly: \texttt{empirical\_type1\_acemoglu.ipynb} pulls a publicly available file from the Dropbox of a study's author, \texttt{empirical\_type1\_ehrlich.ipynb} from an academic's homepage, \texttt{empirical\_type1\_union.ipynb} from Stata Press, and \texttt{empirical\_type2\_mrw.ipynb} comes from \href{https://dataverse.nl}{dataverse.nl}. Two resources come from the \texttt{Python} library \texttt{kagglehub}, and are used in \texttt{empirical\_type4\_airline.ipynb} and \texttt{empirical\_type4\_framingham.ipynb}. Finally, \texttt{empirical\_type5\_gino.ipynb} pulls data from \href{https://researchbox.org}{researchbox.org}. Kindly let us know if you spot that any of the necessary resources become unavailable in their original locations, and we will update our empirical examples in our library accordingly.

\section{Methods}\label{sec:online_methods}
\subsection{Metrics, Methods, and Tests}
\subsubsection{In-sample Metrics}

\texttt{RobustiPy} provides an array of in-sample metrics to evaluate the robustness and quality of analysis. These metrics are calculated for all models and samples, and stored in the results class to maximize their availability to the end-user.

\subsubsubsection*{Coefficient of Determination ($\bar{R}^2$) and Log Likelihoods}

For each ordinary least squares specification, \texttt{RobustiPy} performs the standard -- but adjusted -- in-sample coefficient of determination ($\bar{R}^2$) computation,

% Adjusted R^2 in terms of sums of squares:
\begin{equation}
	\bar{R}^2
	=
	1
	-
	\frac{
		\displaystyle \sum_{i=1}^N \frac{(y_i - \hat y_i)^2  }{(N - P - 1)}
	}{
		\displaystyle \sum_{i=1}^N \frac{(y_i - \bar y)^2}{(N - 1)}
	}
\end{equation}

\noindent where $y_i$ are observed values of each observation $i \in N$, $\hat{y}_i$ are the predicted values, $\bar{y}$ is the mean of the observed data, $N$ is the number of observations for that estimation, and $P$ is the number of parameters. \texttt{RobustiPy} also prints a large number of summary information about all of the calculations which it undertakes, including the maximum value of $\bar{R}^{2}$ and the specification $\pi^*_{\bar{R}^2}$ to which it belongs, as follows:
\begin{equation}
	\pi^*_{\bar{R}^2} = \arg\max_{\pi \in \{1, \dots, |\Pi|\}} \bar{R}^2_\pi \label{r2_max}
\end{equation}

\noindent where $\Pi$ is the set of models in the choice space (see Section \ref{sec:formalisation}) and $\bar{R}^{2}_\pi$ is the $\bar{R}^2$ value of the $\pi$-th model. In the case of logistic regressions, we calculate

\begin{equation}
	R_{\mathrm{McF}}^2
	=
	1
	-
	\frac{
		\displaystyle \sum_{i=1}^N \bigl[y_i\ln\hat y_i \;+\;(1-y_i)\ln(1-\hat y_i)\bigr]
	}{
		\displaystyle \sum_{i=1}^N \bigl[y_i\ln\bar y \;+\;(1-y_i)\ln(1-\bar y)\bigr]
	}\label{eq:mcfadden}
\end{equation}

\noindent which is one minus the ratio of the full-model to null-model log-likelihood, where $\hat y_i$ is the model’s predicted probability that observation $i$ has outcome equal to one. \texttt{RobustiPy} also calculates the specification which maximizes this metric in the Logistic case, equivalent to Eq. \ref{r2_max} (see, for example, \citealp{mcfadden1972conditional} for more details). In the case of the \texttt{LRobust} class, the full model log-likelihood is also returned for each specification.

For \texttt{OLSRobust}, we compute the Gaussian full-model log-likelihood for each specification $\pi$:
\begin{equation}
	\log L_{\pi}^{\mathrm{full}}
	=
	-\frac{N_{\pi}}{2}\bigl(1+\log(2\pi)\bigr)
	-\frac{N_{\pi}}{2}\log\!\bigl(\hat{\sigma}^{2}_{\pi,\mathrm{full}}\bigr),
	\quad
	\hat{\sigma}^{2}_{\pi,\mathrm{full}}
	=
	\frac{1}{N_{\pi}}
	\sum_{i=1}^{N_{\pi}}(y_i-\hat{y}_i)^2.
\end{equation}

\noindent We additionally compute an intercept-only null-model log-likelihood on the same sample:
\begin{equation}
	\log L_{\pi}^{\mathrm{null}}
	=
	-\frac{N_{\pi}}{2}\bigl(1+\log(2\pi)\bigr)
	-\frac{N_{\pi}}{2}\log\!\bigl(\hat{\sigma}^{2}_{\pi,\mathrm{null}}\bigr),
	\quad
	\hat{\sigma}^{2}_{\pi,\mathrm{null}}
	=
	\frac{1}{N_{\pi}}
	\sum_{i=1}^{N_{\pi}}(y_i-\bar{y})^2.
\end{equation}

\noindent For OLS ranking and visualisation, we report the null-relative gain in log-likelihood per observation:
\begin{equation}
	\Delta \log L_{\pi}^{(\mathrm{per\ obs})}
	=
	\frac{
		\log L_{\pi}^{\mathrm{full}}-\log L_{\pi}^{\mathrm{null}}
	}{N_{\pi}}.
\end{equation}

\noindent This metric is invariant to linear rescaling of the dependent variable ($y \mapsto c y$, for $c>0$). The specification $\pi^*_{\Delta \log L}$ which maximizes this OLS metric is presented in the summary method, determined as follows:
\begin{equation}
	\pi^*_{\Delta \log L} = \arg\max_{\pi \in \{1, \dots, |\Pi|\}} \Delta \log L_{\pi}^{(\mathrm{per\ obs})}.
\end{equation}

\noindent where $|\Pi|$ is the total number of specifications in the model space. Raw full-model log-likelihoods are also stored and printed in the summary output.

\subsubsubsection*{Information Criteria}

\texttt{RobustiPy} also computes three information criteria for each specification: Akaike Information Criterion (AIC, see \citealp{akaike1974statistical}), Bayesian Information Criterion (BIC, see \citealp{schwarz1978estimating}), and Hannan-Quinn Information Criterion (HQIC, see \citealp{hannan1979determination}). Each information criterion is calculated from the model's full log-likelihood, as follows:
\begin{equation}
	\textrm{AIC} = 2P - 2 \log L^{\mathrm{full}}
\end{equation}
\begin{equation}\label{eq:bic}
	\textrm{BIC}= P \log N - 2 \log L^{\mathrm{full}}
\end{equation}
\begin{equation}
	\textrm{HQIC} = 2P \log (\log N) - 2 \log L^{\mathrm{full}}
\end{equation}

\noindent where: $P$ is the number of estimated parameters, $N$ is the number of observations, and $\log L^{\mathrm{full}}$ is the full-model log-likelihood. As before, all information criteria are saved in the results class and the minimum value and its associated model is provided by the summary method as before:

\begin{equation}
	\pi^*_{\textrm{IC}} = \arg\min_{\pi \in \{1, \dots,  |\Pi|\}} IC_\pi.
\end{equation}

\noindent where IC $\in$ \{AIC, BIC, HQIC\}. As a rule of thumb, smaller values of any of these indicates a better fit/parameterisation trade-off.

\subsubsection{Out-of-Sample Metrics}\label{methods:oos}

In addition to in-sample metrics, \texttt{RobustiPy} provides multiple k-fold cross-validated out-of-sample metrics for each estimated model. Given a dataset $\mathcal{D}$, defined multiple times by different sets of predictors $\overleftrightarrow{\mathbf{X}}^{(\pi)}$, where $\pi$ indexes different model specifications, we perform analysis with $K$-fold cross-validation to evaluate a predictive model. Each specification and associated dataset $\mathcal{D}^{(\pi)}$ consists of:

\begin{equation}
	\mathcal{D}^{(\pi)} \coloneqq \{(\overleftrightarrow{\mathbf{X}}_i^{(\pi)}, \overleftrightarrow{\mathbf{Z}}_i^{(\pi)}, \overleftrightarrow{\mathbf{Y}}_i^{(\pi)})\}_{i=1}^{N^{(\pi)}},
\end{equation}

\noindent where $N^{(\pi)}$ is the sample size specific to
specification $\pi$ after all preprocessing, $\overleftrightarrow{\mathbf{X}}^{(\pi)}$ represents a specific choice of predictors $\overleftrightarrow{\mathbf{X}}$, $\overleftrightarrow{\mathbf{Z}}^{(\pi)}$ represents a specific choice of control variables $\overleftrightarrow{\mathbf{Z}}$ and $\overleftrightarrow{\mathbf{Y}}^{(\pi)}$ represents a specific choice of dependent variable $\overleftrightarrow{\mathbf{Y}}$. For each specification $\pi$, we divide $\mathcal{D}^{(\pi)}$ into $K$ approximately equal-sized, mutually exclusive subsets $\{ \mathcal{D}_1^{(\pi)}, \mathcal{D}_2^{(\pi)}, \dots, \mathcal{D}_K^{(\pi)} \}$, such that:

\begin{equation}
	\mathcal{D}_k^{(\pi)} \cap \mathcal{D}_j^{(\pi)} = \emptyset \quad \text{for } k \neq j, \quad \bigcup_{k=1}^{K} \mathcal{D}_k^{(\pi)} = \mathcal{D}^{(\pi)}.
\end{equation}

\noindent For ungrouped \texttt{LRobust}, folds are generated with stratified splitting to preserve class balance. When \texttt{group} is provided (for either \texttt{OLSRobust} or \texttt{LRobust}), folds are group-disjoint (\texttt{GroupKFold}), requiring \(K\) to be no larger than the number of unique groups. For ungrouped \texttt{LRobust}, \(K\) must also not exceed the minority-class count after preprocessing. For each fold $k \in \{1, \dots, K\}$ and specification $\pi$, define a training dataset $\mathcal{D}_{\text{train}}^{(k, \pi)} = \mathcal{D}^{(\pi)} \setminus \mathcal{D}_k^{(\pi)}$, and a validation set, $\mathcal{D}_{\text{val}}^{(k, \pi)} = \mathcal{D}_k^{(\pi)}$. Train the model on $\mathcal{D}_{\text{train}}^{(k, \pi)}$ to obtain the learned parameters $\theta^{(k, \pi)}$, and then evaluate it on $\mathcal{D}_{\text{val}}^{(k, \pi)}$. Users then specify or are prompted to select their preferred out-of-sample metric (`\texttt{oos\_metric}') of choice into the \texttt{.fit()} method. For both \texttt{OLSRobust} and \texttt{LRobust}, we calculate the Root Mean Square Error (selected by choosing \texttt{oos\_metric=`rmse'}) for fold $k$ and specification $\pi$, given by:
\begin{equation}
	\text{RMSE}_{k}^{(\pi)} = \sqrt{\frac{1}{|\mathcal{D}_k^{(\pi)}|} \sum_{i \in \mathcal{D}_k^{(\pi)}} \left( y_i - f(\overleftrightarrow{\mathbf{X}}_i^{(\pi)},\overleftrightarrow{\mathbf{Z}}_i^{(\pi)}; \theta^{(k, \pi)}) \right)^2 }.
\end{equation}

We also allow the users to evaluate their out-of-sample model performance with the pseudo-$R^2$ metric as popularised by \cite{salganik2020measuring}, selected through \texttt{oos\_metric=`pseudo-r2'}. This is otherwise known as the Brier Skill Score when evaluating probabilities -- as in the case of \texttt{LRobust} -- and introduced by \cite{brier1950verification}. For fold $k$ and specification $\pi$:

\begin{equation}
	\textrm{Pseudo-R}^{2 (\pi)}_{\space k} = 1 - \frac{\sum_{i \in \mathcal{D}_k^{(\pi)}} \left( y_i - f(\overleftrightarrow{\mathbf{X}}_i^{(\pi)},\overleftrightarrow{\mathbf{Z}}_i^{(\pi)}; \theta^{(k, \pi)}) \right)^2 }{\sum_{i \in \mathcal{D}_k^{(\pi)}} \left( y_i - \bar{y}^{(\pi)} \right)^2 },
\end{equation}

\noindent where $\bar{y}^{(\pi)}$ is the mean of $y_i$ in the training set of each fold. In the case of \texttt{LRobust} when calculating an out-of-sample R$^2$, we also allow the user to choose to evaluate their model with McFadden's R$^2$-metric (\texttt{oos\_metric{=}`mcfaddens\allowbreak-r2'}), but instead use the mean of the training dataset in the calculation of the denominator of Eq. \ref{eq:mcfadden}. %\noindent For logistic regression models, we also compute McFadden’s Pseudo-$R^2$, defined for fold $j$ and specification $s$ as:
%\begin{equation}
%R^2_{\text{McF}, j, s} = 1 - \frac{\log L_{\text{full}}^{(j, s)}}{\log L_{\text{null}}^{(j, s)}},
%\end{equation}
%where $\log L_{\text{full}}^{(j, s)}$ is the log-likelihood of the fitted model evaluated on the validation set, and $\log L_{\text{null}}^{(j, s)}$ is the log-likelihood of a null model with only an intercept. This provides a normalized measure of improvement over the baseline.
%

We additionally include the Cross Entropy (also referred to as Log Loss) as an out-of-sample metric for evaluating probabilistic predictions in binary classification models (\texttt{oos\_metric=`cross-entropy'}). This metric penalises confident but incorrect predictions more heavily, making it particularly useful for assessing well-calibrated models. For fold $k$ and specification $\pi$, the empirical cross entropy is defined as:

\begin{equation}
	\begin{split}
		\text{Empirical Cross Entropy}_{k}^{(\pi)}
		 & = -\frac{1}{|\mathcal{D}_k^{(\pi)}|}
		\sum_{i \in \mathcal{D}_k^{(\pi)}} \Big[
		y_i \log\!\big(
		f(\overleftrightarrow{\mathbf{X}}_i^{(\pi)},
		\overleftrightarrow{\mathbf{Z}}_i^{(\pi)};
		\theta^{(k, \pi)})
		\big)                                   \\
		 & \qquad\qquad\qquad\quad
		+ (1 - y_i)\,
		\log\!\big(
		1 - f(\overleftrightarrow{\mathbf{X}}_i^{(\pi)},
		\overleftrightarrow{\mathbf{Z}}_i^{(\pi)};
		\theta^{(k, \pi)})
		\big)
		\Big].
	\end{split}
\end{equation}

\noindent In addition, we include the InterModel Vigorish (IMV) --  recently introduced by \cite{domingue2025intermodel}, \cite{domingue2024intermodel}, and \cite{Zhang2023IMV} -- for binary classifications (\texttt{oos\_metric=`imv'}). For fold $k$ and specification $\pi$, it is defined as:

\begin{equation}
	\text{IMV}_{k}^{(\pi)} = \frac{w_{\text{enhanced}}^{(k,\pi)} - w_{\text{null}}^{(k,\pi)}}{w_{\text{null}}^{(k,\pi)}},
\end{equation}
where $w_{\text{null}}^{(k,\pi)}$ and $w_{\text{enhanced}}^{(k,\pi)}$ are entropic representations of model fit. Each $w$ is the value minimising the difference between $w \log w - (1 - w) \log (1 - w) $ and $ \frac{1}{|\mathcal{D}_k^{(\pi)}|}\sum_{i\in \mathcal{D}_k^{(\pi)}}y_i\log \hat{y}_i+(1-y_i)\log (1-\hat{y}_i)$. The $\hat{y}_i$ is a predicted value from the model fit with choice $\pi$ for $w_{\text{enhanced}}^{(k,\pi)}$, and for $w_{\text{null}}^{(k,\pi)}$ as $\bar{y}_i$ in the validation set $\mathcal{D}_k^{(\pi)}$. These metrics are then averaged over the corresponding $K$ folds, as
%The final RMSE for specification $s$ across all folds is computed as:
%\begin{equation}
$\text{RMSE}_{\text{CV}, \pi} = \frac{1}{K} \sum_{k=1}^{K} \text{RMSE}_{k,\pi}$,
%\end{equation}
%
%The final $R^2$ for specification $s$ across all folds is computed as:
%\begin{equation}
% ← preceding comma in running text
$\text{Pseudo-R}^{2}_{\text{CV},\pi}
	= \frac{1}{K}\sum_{k=1}^{K}\text{Pseudo-}R^{2}_{k,\pi}$,
$\text{McFadden's-R}^{2}_{\text{CV},\pi}
	= \frac{1}{K}\sum_{k=1}^{K}\text{McFadden's-R}^{2}_{k,\pi}$, $\text{Cross Entropy}_{\text{CV}, \pi} = \frac{1}{K}\sum_{k=1}^{K} \text{Cross Entropy}_{k,\pi}$
and $\text{IMV}_{\text{CV},\pi}
	\;=\; \frac{1}{K}\sum_{k=1}^{K}\text{IMV}_{k,\pi}$.

%\end{equation}
To compare multiple specifications, we compute, store, and make readily available the metrics for each $\pi$ and analyse their distributions across arrays stored as follows:
\begin{equation}
	\mathcal{S}_{\text{RMSE}} \coloneqq [ \text{RMSE}_{\text{CV}, \pi} ]_{\pi=1}^{|\Pi|},
\end{equation}
\begin{equation}
	\mathcal{S}_{\text{Pseudo-R}^2} \coloneqq [ \text{Pseudo-R}^2_{\text{CV}, \pi} ]_{\pi=1}^{|\Pi|},
\end{equation}
\begin{equation}
	\mathcal{S}_{\text{McFadden's-R}^2} \coloneqq [\text{McFadden's-R}^2_{\text{CV}, \pi} ]_{\pi=1}^{|\Pi|},
\end{equation}
\begin{equation}
	\mathcal{S}_{\text{Cross Entropy}} \coloneqq [\text{Cross Entropy}_{\text{CV}, \pi}]_{\pi=1}^{|\Pi|},
\end{equation}
\begin{equation}
	\mathcal{S}_{\text{IMV}} \coloneqq [  \text{IMV}_{\text{CV}, \pi} ]_{\pi=1}^{|\Pi|},
\end{equation}

\noindent where $|\Pi|$ is the total number of models estimated. %\par
%\texttt{RobustiPy} will store the sets $\mathcal{S}_{\text{RMSE}}$ and $\mathcal{S}_{R^2}$ in the $results$ class and it will deliver the minimum, maximum, median and mean of the metrics across all specification in the $summary$ method.

\subsubsection{Computational Abilities}
%In addition to the specification curve itself,
\texttt{RobustiPy} performs an array of computations for analysing each model in the choice space ($\pi \in \Pi$).

\subsubsubsection*{Bootstrapping}
To quantify uncertainty in the regression estimates in the specification/choice curve, \texttt{RobustiPy} performs a bootstrap procedure to estimate the confidence interval of the effect between $\overleftrightarrow{\mathbf{Y}}$ and the key estimand of interest, $\mathbf{X}_{1}$. For each choice $\pi\in \Pi$:

\begin{enumerate}
	\item Fit the model to observed data
	      \begin{equation}
		      \hat\omega_{\pi} = F(\overleftrightarrow{\mathbf{Y}}_{\pi}, \overleftrightarrow{\mathbf{X}}_{\pi}, \overleftrightarrow{\mathbf{Z}_{\pi}})
	      \end{equation}
	      where $\hat\omega_{\pi}$ is typically the coefficient of the predictor of interest $\mathbf{X}_{1,\pi}$
	\item Draw \(B\) bootstrap replicates \(b\) from \(\mathcal{D}^{(\pi)}\). For ungrouped models, rows are resampled with replacement; when \texttt{group} is provided, groups are resampled with replacement (cluster bootstrap) until the requested row target (\texttt{sample\_size} -- defaulting to full sample size) is reached -- so realized replicate size may vary.
	\item Re-estimate the model in each replicate $b$
	      \begin{equation}
		      \hat\omega_{\pi}^{(b)} = F(\overleftrightarrow{\mathbf{Y}}_{\pi}^{(b)}, \overleftrightarrow{\mathbf{X}}_{\pi}^{(b)}, \overleftrightarrow{\mathbf{Z}_{\pi}}^{(b)}) \qquad
		      b = 1,\dots,B.
	      \end{equation}
	\item Construct the empirical distribution of coefficients $\hat\omega_{\pi}^{(b)}$ across all bootstrap replicates
	      \begin{equation}
		      \mathcal{S}_{\pi}^{\text{boot}} \coloneqq \{ \hat\omega_{\pi}^{(b)}  \}_{b=1}^{B}.
	      \end{equation}
\end{enumerate}

\noindent Let $c \in (0,1)$ denote the user-specified confidence level (\texttt{ci}=$c$, default $c=0.95$). The corresponding confidence interval is:
\begin{equation}
	\text{CI}_{\omega} = \left\{\left[ Q_{(1-c)/2}(\mathcal{S}_{\pi}^{\text{boot}}), \quad Q_{1-(1-c)/2}(\mathcal{S}_{\pi}^{\text{boot}}) \right] \colon \pi \in \Pi \right\}
\end{equation}
\noindent where $Q_{\alpha}$ denotes the empirical quantile function. Values (for the specification curve subfigure) can be optionally LOESS-smoothed for better visualization, but raw estimates are kept. If singleton filtering yields unstable grouped samples, implementation may fall back to unfiltered grouped draws to avoid hard failure.

\subsubsection*{Bayesian Model Averaging for control coefficients (BIC-implied priors)}

\texttt{RobustiPy} performs Bayesian Model Averaging (BMA) for variables specified in \(\overleftrightarrow{\mathbf Z}_{\pi}\). The focal predictor(s) in \(\overleftrightarrow{\mathbf X}_{\pi}\) are included in every specification, so they are not averaged over. For clarity, the equations below follow one generic control variable (indexed as \(p\)), where the same steps are repeated for every other control;

\begin{equation}
	\overleftrightarrow{\mathbf Y}_{\pi}
	=
	\bigl[
		\overleftrightarrow{\mathbf X}_{\pi},
		\overleftrightarrow{\mathbf Z}_{\pi}
		\bigr]
	\boldsymbol\beta_{\pi}
	+\boldsymbol\epsilon_{\pi},
	\qquad
	\boldsymbol\epsilon_{\pi}\sim
	\mathcal N\!\bigl(\mathbf 0,\,
	\sigma_{\pi}^{2}\mathbf I_{N_\pi}\bigr).
\end{equation}

\noindent The \(p^{\text{th}}\) coefficient in \(\boldsymbol\beta_{\pi}\) is denoted
\begin{equation}
	\omega_{p,\pi}\equiv\beta_{p,\pi}.
\end{equation}

\noindent With a uniform prior over models, the posterior weight of model \(M_\pi\) is
\begin{equation}
	P(M_\pi\mid\mathcal D)
	=
	\frac{\exp\!\bigl[-\tfrac12\,\text{BIC}_{\pi}\bigr]}
	{\displaystyle\sum_{\pi'\in\Pi}
		\exp\!\bigl[-\tfrac12\,\text{BIC}_{\pi'}\bigr]},
\end{equation}
where
\begin{equation}\label{eq:bic_pi}
	\text{BIC}_{\pi}
	=
	P_{\pi}\,\log N_{\pi}
	\;-\;
	2\,\log\hat L_{\pi},
	\qquad\text{(cf.\ Eq.~\ref{eq:bic})}
\end{equation}

\noindent Let \(\hat\omega_{p,\pi}\) be the within-model posterior mean (equal to the OLS or Maximum Likelihood based estimate under a vague prior). The BMA estimate for index \(p\) is
\begin{equation}
	\mathbb E[\omega_{p}\mid\mathcal D]
	=
	\sum_{\pi\in\Pi}
	P(M_\pi\mid\mathcal D)\,
	\hat\omega_{p,\pi}.
\end{equation}

\noindent Repeating this step for every \(p=1,\dots,P\) yields the full vector
\begin{equation}
	\widehat{\boldsymbol\omega}^{\text{BMA}}
	=
	\bigl(
	\mathbb E[\omega_{1}\mid\mathcal D],
	\dots,
	\mathbb E[\omega_{p}\mid\mathcal D]
	\bigr)^{\!\top}.
\end{equation}

\subsubsection*{SHAP Values for the Full Linear–Additive Model}

\texttt{RobustiPy} computes SHapley Additive exPlanations (SHAP) for the single `full' specification containing all predictors and controls. After dropping missing values, the data are split into an 80\% training set and a 20\% held-out test set. For \texttt{OLSRobust} with \texttt{group}, demeaning is applied before fitting; for \texttt{LRobust}, logistic fixed-effects demeaning is not applied and \texttt{group} is used for grouped cross-validation/bootstrap resampling.  We fit ordinary least squares (or a Logistic Regression with an L2 penalty and inverse regularisation strength of 0.1, depending on the class used) on the training data:

\begin{equation}
	\widehat{Y}_{i}
	=
	[\overleftrightarrow{\mathbf X}_{i},\overleftrightarrow{\mathbf Z}_{i}]\;\widehat{\boldsymbol\beta},
\end{equation}
\noindent for $i=1,\dots,N_{\mathrm{test}}$ where $\overleftrightarrow{\mathbf X}_{i}$ and $\overleftrightarrow{\mathbf Z}_{i}$ are row‐vectors of the test‐set features (including the intercept). Let
\begin{equation}
	\mu_{p}
	=
	\mathbb{E}_{\mathrm{train}}[X_{p}]
\end{equation}
and
\begin{equation}
	v(\emptyset)
	=
	\sum_{p=1}^{P}\widehat\beta_{p}\,\mu_{p}.
\end{equation}
Under linearity, the exact SHAP values are
\begin{equation}
	\widehat\phi_{p,i}
	=
	\widehat\beta_{p}\,\bigl(X_{p,i} - \mu_{p}\bigr),
\end{equation}
for $p=1,\dots,P$ and $i=1,\dots,N_{\mathrm{test}}$,
which satisfy
\begin{equation}
	\widehat{Y}_{i}
	=
	v(\emptyset)
	+
	\sum_{p=1}^{P}\widehat\phi_{p,i}.
\end{equation}

\noindent Rather than aggregating them internally, \texttt{RobustiPy} simply returns
\begin{equation}
	\texttt{shap\_return}
	=
	\Bigl(
	[\widehat\phi_{p,i}]_{i=1,\dots,N_{\mathrm{test}};\,p=1,\dots,P},
	\;\mathbf X^{\mathrm{test}}
	\Bigr).
\end{equation}
SHAP values quantify how much each predictor raises or lowers the model’s fitted outcome for a given observation; larger absolute SHAP values indicate stronger, direction-specific influence on that prediction. Users may then compute summaries (e.g.\ mean absolute SHAP per feature) or visualize these values as desired.

\subsubsubsection*{Specification–curve inference}

Beyond per–specification uncertainty, \texttt{RobustiPy} implements a
\emph{curve-level} test of the sharp null
\[
	H_{0} : \beta_{1} = 0,
	\qquad
	H_{1} : \beta_{1} \neq 0.
\]
asking whether the entire pattern of estimates across the specification space $\Pi$ could plausibly arise if the treatment effect were truly zero. The procedure mirrors the outcome bootstrap introduced earlier, but
performs the aggregation \emph{after} estimation.

\begin{enumerate}

	\item \textbf{Full-sample fits}:
	      For every specification $\pi\in\Pi$, fit on its observed data
	      $\mathcal D^{(\pi)}$, storing the coefficient and $p$-value
	      $(\hat\omega_{\pi}, p_{\pi})$.

	\item \textbf{Outcome bootstrap (sampling variability)}:
	      Draw $B$ i.i.d.\ bootstrap resamples of analysis units—\emph{shared across all
		      specifications}. If no grouping variable is supplied, units are row indices.
	      If a grouping variable is supplied, units are unique group identifiers
	      (cluster bootstrap with replacement, retaining all rows from sampled groups,
	      including multiplicity). Re-estimate $\hat\omega_{\pi}^{(b)}$ for each
	      $\pi\in\Pi$ and $b=1,\dots,B$. This yields the finite-sample distribution of each
	      summary statistic under the empirical design.

	\item \textbf{Null bootstrap ($y^\star$ construction)}:
	      Enforce the sharp null by subtracting off the estimated treatment effect
	      \begin{equation}
		      Y^{\star}_{\pi} \;=\; Y_{\pi} - \mathbf X_{1,\pi}\,\hat\omega_{\pi},
	      \end{equation}
	      and re–fit each specification on $Y^{\star}_{\pi}$ using the same $B$ bootstrap resamples at the same sampling level (row-level when ungrouped; group-level when grouped). This generates the joint null distribution of curve–level functionals
	      while preserving the correlation structure induced by resampling the same analysis units across specifications (rows when ungrouped, groups when grouped).

	\item \textbf{Curve-level functionals}:
	      For a vector $A=\{a_{\pi}\}_{\pi\in\Pi}$ and (optionally) $p$-values
	      $P=\{p_{\pi}\}_{\pi\in\Pi}$, define
	      \begin{align}
		      S_{1}(A) & \coloneqq \operatorname{median}_{\pi} a_{\pi},
	      \end{align}
	      \begin{align}
		      S_{2}(A) & \coloneqq \min_{\pi} a_{\pi},
	      \end{align}
	      \begin{align}
		      S_{3}(A) & \coloneqq \max_{\pi} a_{\pi},
	      \end{align}
	      \begin{align}
		      S_{4}(A) & \coloneqq \sum_{\pi}\mathbbm{1}\{a_{\pi}>0\},
	      \end{align}
	      \begin{align}
		      S_{5}(A) & \coloneqq \sum_{\pi}\mathbbm{1}\{a_{\pi}<0\},
	      \end{align}
	      \begin{align}
		      S_{6}(A,P) & \coloneqq \sum_{\pi}\mathbbm{1}\{p_{\pi}<0.05\},
	      \end{align}
	      \begin{align}
		      S_{7}(A,P) & \coloneqq \sum_{\pi}\mathbbm{1}\{a_{\pi}>0\}\,\mathbbm{1}\{p_{\pi}<0.05\},
	      \end{align}
	      \begin{align}
		      S_{8}(A,P) & \coloneqq \sum_{\pi}\mathbbm{1}\{a_{\pi}<0\}\,\mathbbm{1}\{p_{\pi}<0.05\}.
	      \end{align}

	\item \textbf{Two–sided null $p$–values}:
	      For each functional $S$, let $S_{\text{obs}}$ denote the value on the
	      observed coefficients $\hat\omega_{\pi}$, and
	      $\{S^{\star(b)}\}_{b=1}^{B}$ its values under null bootstraps.
	      The exact Monte Carlo $p$–value is

	      \begin{equation}\label{eq:two-sided-p}
		      p(S_{\text{obs}})=
		      2\min\!\left\{
		      \frac{1+\sum_{b=1}^{B}\mathbbm{1}\!\left\{S^{\star(b)}\ge S_{\text{obs}}\right\}}{B+1},\,
		      \frac{1+\sum_{b=1}^{B}\mathbbm{1}\!\left\{S^{\star(b)}\le S_{\text{obs}}\right\}}{B+1}
		      \right\},
		      \qquad
		      p(S_{\text{obs}})\leftarrow \min\{1,\;p(S_{\text{obs}})\}.
	      \end{equation}

	      clipped at $1$.
	      This ensures two–sided inference: extremeness in either tail counts
	      against $H_{0}$.

	\item \textbf{Model–class caveat}:
	      Eq.~\ref{eq:two-sided-p} is formally justified under the ideal i.i.d.\ linear model assumptions of OLS, under which the row bootstrap consistently approximates the sampling distribution of smooth functionals of the estimator. For grouped models, \texttt{RobustiPy} uses cluster-level resampling based on the supplied grouping variable; inference is therefore conditional on the supplied grouping variable correctly capturing the relevant dependence structure, and for logit models the score is heteroskedastic and the outcome bounded, so naive resampling may not reproduce the correct finite-sample law. In these settings the resulting null $p$–values may be miscalibrated and should therefore be interpreted as descriptive diagnostics rather than exact hypothesis tests. If curve-level summaries are computed using data-dependent weights (e.g.\ information-criterion weights), those weights must be recomputed within each bootstrap replicate to preserve coherence between the observed statistic and its null reference distribution. Otherwise the induced null distribution need not correspond to the reported weighted functional.

	\item \textbf{Meta--analytic Stouffer test (null--calibrated)}:
	      Let $\Pi$ denote the specification set. For each $\pi\in\Pi$ we retain the
	      full--sample estimate and $p$--value $(\hat\beta_\pi,\,p_\pi)$. When joint
	      inference is enabled, the null bootstrap additionally provides draws
	      $\{(\hat\beta_\pi^{\star(b)},\,p_\pi^{\star(b)})\}_{b=1}^B$ computed on the
	      pseudo--outcome $Y_\pi^\star$ and using the same shared analysis unit resamples across specifications.

	      \textit{Clipping and signed $z$--scores.}
	      Define the numerically clipped $p$--values
	      \begin{equation}\label{eq:stouffer-clip}
		      p_{\pi}^{\mathrm{clip}}
		      \;\coloneqq\;
		      \min\!\Bigl(\max(p_{\pi},\;\varepsilon),\;1-\delta\Bigr),
		      \qquad
		      \varepsilon \;\coloneqq\; \max\!\bigl(10^{-300},\,\tau_{\mathrm{mach}}\bigr),
		      \ \ \delta \;\coloneqq\; 10^{-16},
	      \end{equation}
	      where $\tau_{\mathrm{mach}}$ denotes the smallest strictly positive
	      \emph{representable} double--precision number.\footnote{In code this is
		      \texttt{np.finfo(float).tiny}.}
	      The observed signed $z$--scores are
	      \begin{equation}\label{eq:stouffer-zobs}
		      z_\pi
		      \;\coloneqq\;
		      \operatorname{sign}(\hat\beta_\pi)\,
		      \Phi^{-1}\!\left(1-\tfrac{p_{\pi}^{\mathrm{clip}}}{2}\right),
	      \end{equation}
	      with $\Phi^{-1}$ the standard normal quantile. As a safeguard, any
	      non--finite values (which can occur only at extreme tails) are clipped to
	      the range $[-37,37]$.

	      \textit{Null $z$--scores and screening.}
	      From the null draws define, for each $b=1,\dots,B$,
	      \begin{equation}\label{eq:stouffer-znull}
		      z_{\pi}^{\star(b)}
		      \;\coloneqq\;
		      \operatorname{sign}(\hat\beta_\pi^{\star(b)})\,
		      \Phi^{-1}\!\left(1-\tfrac{(p_{\pi}^{\star(b)})^{\mathrm{clip}}}{2}\right).
	      \end{equation}
	      Let $Z^{\star(b)}_{\mathrm{vec}} \coloneqq (z_{\pi}^{\star(b)})_{\pi\in\Pi}$.
	      Specifications with (near) zero null variability are excluded prior to dependence estimation, i.e.\ we retain
	      \begin{equation}
		      \Pi_{\mathrm{keep}}
		      \;\coloneqq\;
		      \Bigl\{\pi\in\Pi:\ \widehat{\operatorname{sd}}\bigl(z_{\pi}^{\star(1:B)}\bigr)>10^{-14}\Bigr\},
	      \end{equation}
	      and subsequently work with the restricted vectors
	      $Z^{\star(b)}_{\mathrm{vec,keep}} \coloneqq (z_{\pi}^{\star(b)})_{\pi\in\Pi_{\mathrm{keep}}}$.

	      \textit{Dependence correction from the null bootstrap.}
	      Using the $B$ null vectors, form the empirical correlation matrix
	      \begin{equation}
		      \widehat R \;\coloneqq\; \operatorname{Corr}\!\Bigl(Z^{\star(1)}_{\mathrm{vec,keep}},\dots,Z^{\star(B)}_{\mathrm{vec,keep}}\Bigr).
	      \end{equation}
	      To ensure numerical stability, $\widehat R$ is projected onto the cone of
	      positive semidefinite matrices (eigenvalue clipping), symmetrized, forced
	      to have unit diagonal, and ridge--regularized:
	      \begin{equation}\label{eq:stouffer-sigmahat}
		      \widehat\Sigma
		      \;\coloneqq\;
		      \widehat R_{\mathrm{psd}} \;+\; 10^{-12} I.
	      \end{equation}
	      If $\widehat\Sigma$ cannot be constructed (e.g.\ too few null draws), the
	      implementation falls back to the independence denominator.

	      \textit{Combined statistic.}
	      With nonnegative weights $w_\pi$ (default $w_\pi\equiv 1$) for
	      $\pi\in\Pi_{\mathrm{keep}}$, define
	      \begin{equation}\label{eq:stouffer-Zobs}
		      Z_{\mathrm{obs}}
		      \;\coloneqq\;
		      \frac{\sum_{\pi\in\Pi_{\mathrm{keep}}} w_\pi z_\pi}{\sqrt{w^\top \widehat\Sigma\, w}},
	      \end{equation}
	      where, under the independence fallback, the denominator is
	      $\sqrt{\sum_{\pi\in\Pi_{\mathrm{keep}}} w_\pi^2}$.

	      \textit{Null calibration (as implemented).}
	      For each null draw $b$, compute the corresponding combined statistic using
	      the \emph{same} denominator as in~\eqref{eq:stouffer-Zobs},
	      \begin{equation}\label{eq:stouffer-Znull}
		      Z^{\star(b)}
		      \;\coloneqq\;
		      \frac{\sum_{\pi\in\Pi_{\mathrm{keep}}} w_\pi z_{\pi}^{\star(b)}}{\sqrt{w^\top \widehat\Sigma\, w}},
		      \qquad b=1,\dots,B.
	      \end{equation}
	      The two--sided null--calibrated Monte Carlo $p$--value is
	      \begin{equation}\label{eq:stouffer-p}
		      p_{\mathrm{Stouffer}}
		      \;\coloneqq\;
		      \frac{1+\sum_{b=1}^B \mathbbm{1}\!\left\{|Z^{\star(b)}|\ge |Z_{\mathrm{obs}}|\right\}}{B+1}.
	      \end{equation}

	      \textit{Interpretation.}
	      Large positive $Z_{\mathrm{obs}}$ indicates broadly concordant evidence for
	      $\beta_1>0$ across specifications; large negative $Z_{\mathrm{obs}}$ indicates
	      concordant evidence for $\beta_1<0$; values near zero indicate weak or
	      conflicting evidence. Invalid or missing $(\hat\beta_\pi,p_\pi)$ are omitted
	      from $\Pi_{\mathrm{keep}}$, and dependence is estimated from the null bootstrap
	      when available.

	\item \textbf{Stored outputs}:
	      All curve–level summaries $S_{k}$, their null $p$–values, and the
	      $(Z,p_{\text{Stouffer}})$ pair are saved in
	      \verb|OLSResult.inference| for downstream use in
	      \verb|.summary()| and in visualization methods.
\end{enumerate}

\subsubsubsection*{Sub-sampling large specification spaces}

When the number of admissible outcome composites
\(\Pi_{\overleftrightarrow{\mathbf Y}}\) or covariate sets
\(\Pi_{\overleftrightarrow{\mathbf Z}}\) is very large,
estimating \emph{every} model in
\(\Pi=\Pi_{\overleftrightarrow{\mathbf Y}}\!\times\!
\Pi_{\overleftrightarrow{\mathbf X}}\!\times\!
\Pi_{\overleftrightarrow{\mathbf Z}}\!\times\!
\Pi_{\overleftrightarrow{F}}\)
can be computationally prohibitive.
\texttt{RobustiPy} therefore lets the user draw a random subset of
specifications before any fitting begins.

\begin{enumerate}

	\item \textbf{Determine target sizes.}
	      Choose integers
	      \(m_{Y}\le |\Pi_{\overleftrightarrow{\mathbf Y}}|\) and
	      \(m_{Z}\le |\Pi_{\overleftrightarrow{\mathbf Z}}|\),
	      the number of outcome– and covariate–options you wish to keep
	      (defaults are the full sizes).

	\item \textbf{Uniform sub-sampling.}
	      Independently sample, \emph{without replacement},
	      \begin{equation}
		      \tilde{\Pi}_{\overleftrightarrow{\mathbf Y}}
		      \sim \mathrm{Unif}\!\left(
		      \left\{
		      A \subset \Pi_{\overleftrightarrow{\mathbf Y}}
		      :\; |A| = m_Y
		      \right\}
		      \right),
	      \end{equation}

	      \begin{equation}
		      \tilde{\Pi}_{\overleftrightarrow{\mathbf Z}}
		      \sim \mathrm{Unif}\!\left(
		      \left\{
		      A \subset \Pi_{\overleftrightarrow{\mathbf Z}}
		      :\; |A| = m_Z
		      \right\}
		      \right),
	      \end{equation}
	      using NumPy’s fast random sampler.

	\item \textbf{Construct the reduced model space.}
	      The analysis then proceeds on
	      \begin{equation}
		      \tilde{\Pi}
		      \;=\;
		      \tilde{\Pi}_{\overleftrightarrow{\mathbf Y}}
		      \times
		      \Pi_{\overleftrightarrow{\mathbf X}}
		      \times
		      \tilde{\Pi}_{\overleftrightarrow{\mathbf Z}}
		      \times
		      \Pi_{\overleftrightarrow{F}},
	      \end{equation}
	      which typically cuts runtime by orders of magnitude while still
	      giving a representative picture of the full specification curve.
\end{enumerate}

\noindent In the current release, pre-fit sub-sampling is fully implemented in \texttt{OLSRobust} via \texttt{composite\_sample} and \texttt{z\_specs\_sample\_size}; \texttt{LRobust} does not yet expose fully symmetric controls across both outcome- and covariate-space sampling. The random seeds are user-controllable, ensuring that
sub-samples—and therefore results—are fully replicable.

\newpage
\bibliography{references}%References – we do not place limits on reference %lists.
\bibliographystyle{chicago}

\clearpage
\section*{Supplementary Information}
\subsection*{Supplementary Figures}

% Reset and redefine the figure (and table) numbering
\setcounter{figure}{0}                              % reset figure counter
\renewcommand{\thefigure}{S\arabic{figure}}         % prefix “S” to figures

\setcounter{table}{0}                               % (optional) reset table counter
\renewcommand{\thetable}{S\arabic{table}}           % (optional) prefix “S” to tables
\setcounter{codeblock}{0}
\renewcommand{\theHcodeblock}{supp.\arabic{codeblock}}

\begin{figure}[ht]
	\centering
	\includegraphics[width=\textwidth]{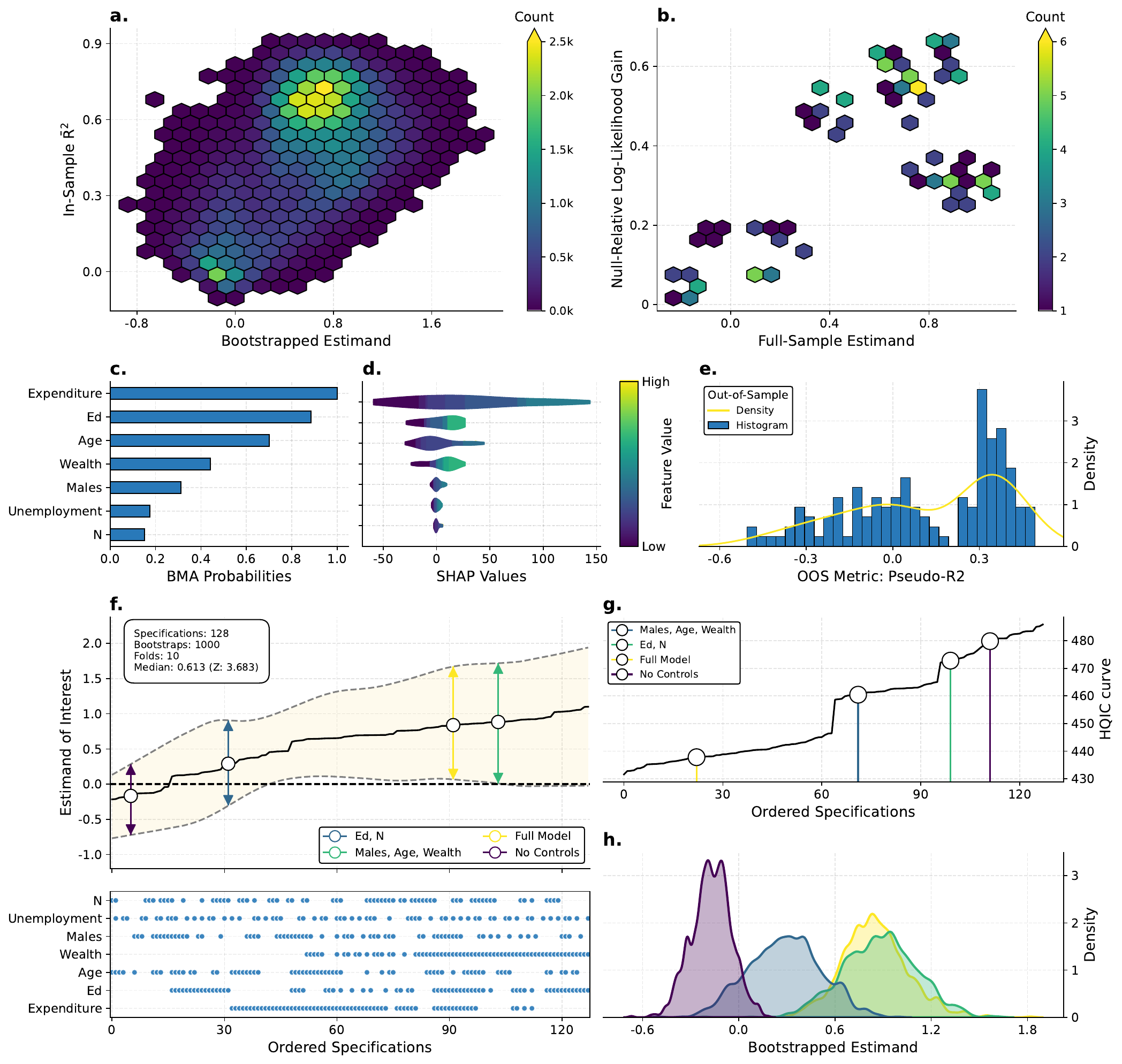}
	\caption{An application of \texttt{RobustiPy} to criminology: \cite{ehrlich1973participation} revisited, based on a revised dataset from \cite{vandaele1987participation}. The dependent variable is state-level crime rate, while the estimand of interest is income inequality.}
	\label{fig:S1}
\end{figure}

\clearpage

\begin{figure}[ht]
	\centering
	\includegraphics[width=\textwidth]{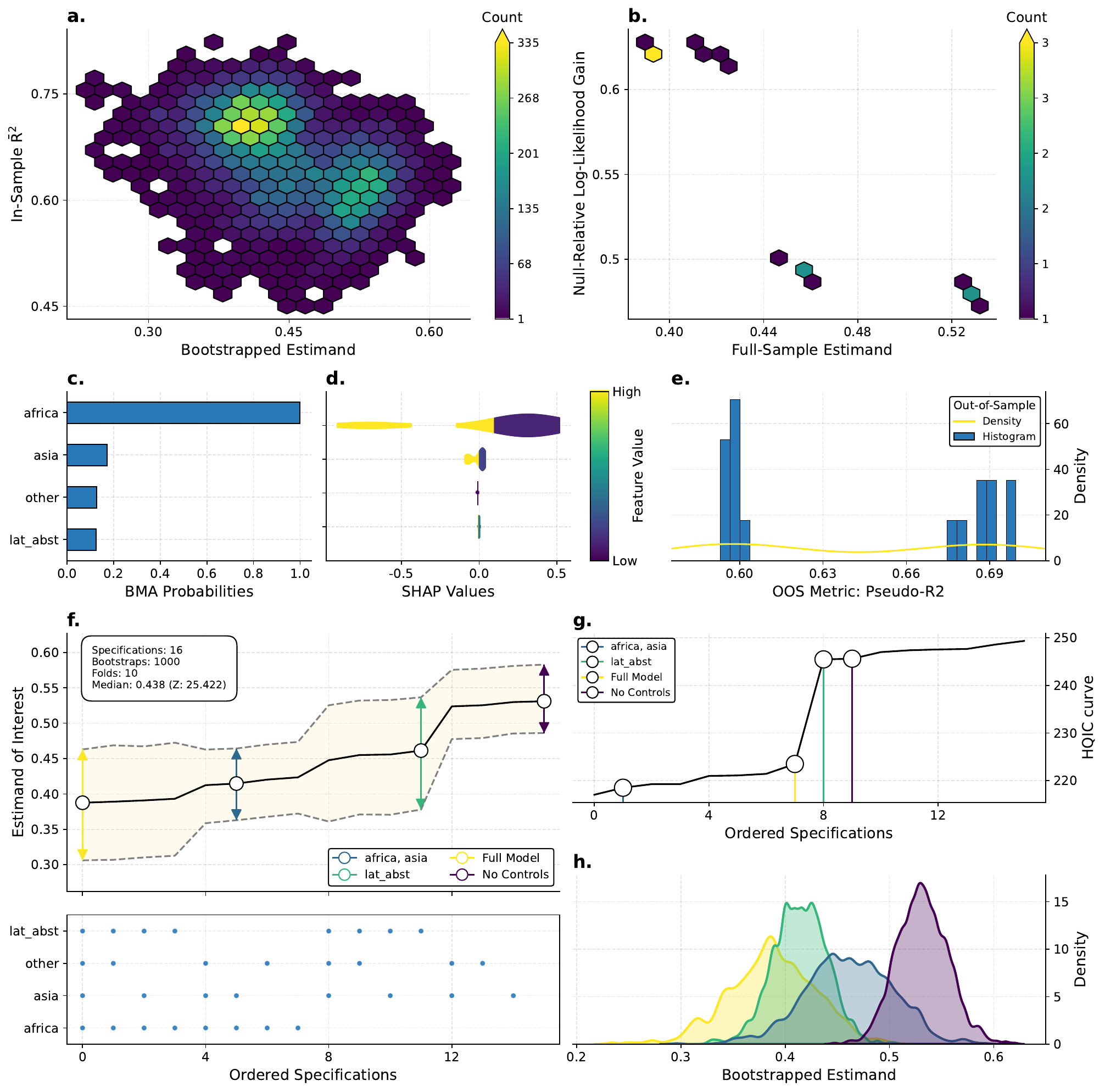}
	\caption{A re-examination of the influential \cite{acemoglu2001colonial} paper. Our analysis spans from the core, baseline OLS model, to the fullest OLS model as specified in Table 2 of the aforementioned paper, where the variation of all intermediate specifications are analysed by \texttt{RobustiPy}. The baseline models without any and with all controls are highlighted in subfigures `\textbf{f.}', `\textbf{g.}' and `\textbf{h.}'.}
	\label{fig:S2}
\end{figure}
\clearpage

\begin{figure}[ht]
	\centering
	\includegraphics[width=\textwidth]{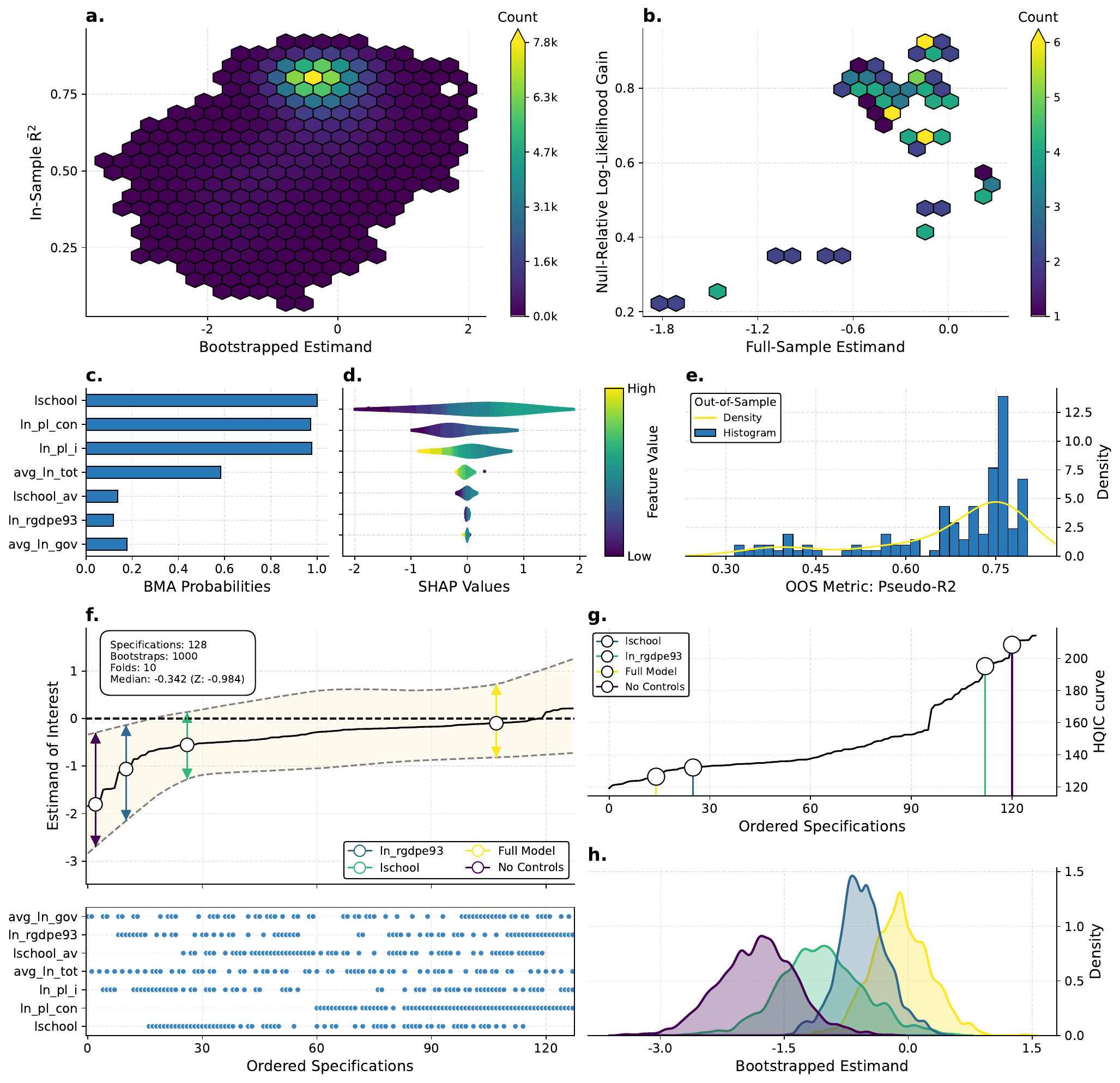}
	\caption{A re-analysis of the Solow Growth Model as per \cite{mankiw1992contribution}. With this instantiation of \texttt{RobustiPy}, we order the logarithm of (n+g+$\delta$) -- averaged over the preceding years and empirically calibrated for all countries -- first in our $\mathrm{x}$ array.}
	\label{fig:S3}
\end{figure}

\clearpage

\begin{figure}[ht]
	\centering
	\includegraphics[width=\textwidth]{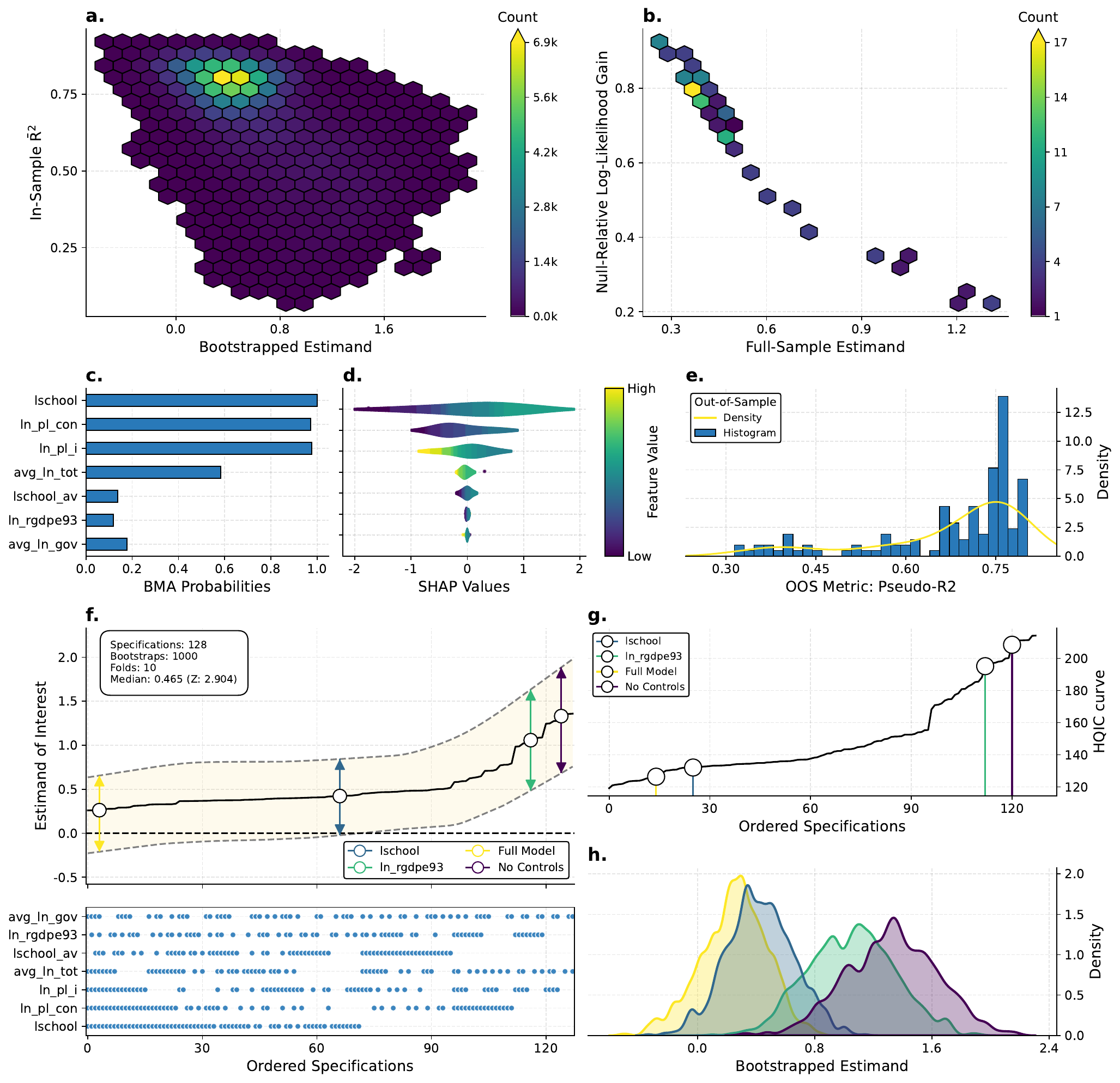}
	\caption{A re-analysis of the Solow Growth Model as per \cite{mankiw1992contribution}. With this instantiation of \texttt{RobustiPy}, we order the logarithm of $\frac{I}{Y}$ -- averaged over the preceding years and empirically calibrated for all countries -- first in our $\mathrm{x}$ array.}
	\label{fig:S4}
\end{figure}

\clearpage

\begin{figure}[ht]
	\centering
	\includegraphics[width=\textwidth]{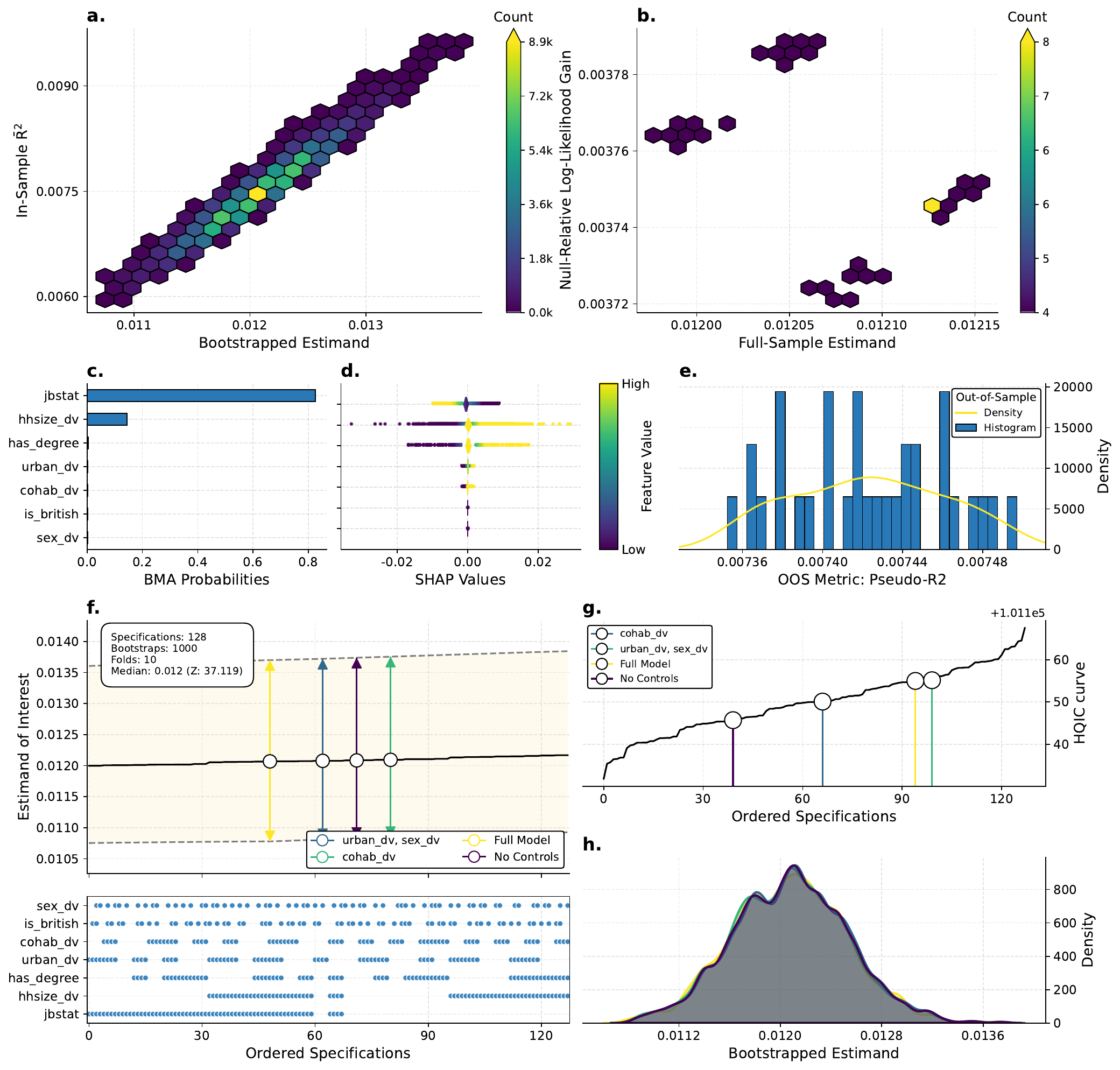}
	\caption{\textbf{Example output of \texttt{RobustiPy} utilising the UKHLS dataset}. The dependent variable is a binary representation of whether a survey recipient is in `good health', estimated as a linear probability model. This example showcases individuals grouped longitudinally via the \texttt{group} input to the `\texttt{.fit()}' method (in this case, with \texttt{group=`pipd'}).}
	\label{fig:S5}
\end{figure}

\clearpage

\begin{figure}[ht]
	\centering
	\includegraphics[width=\textwidth]{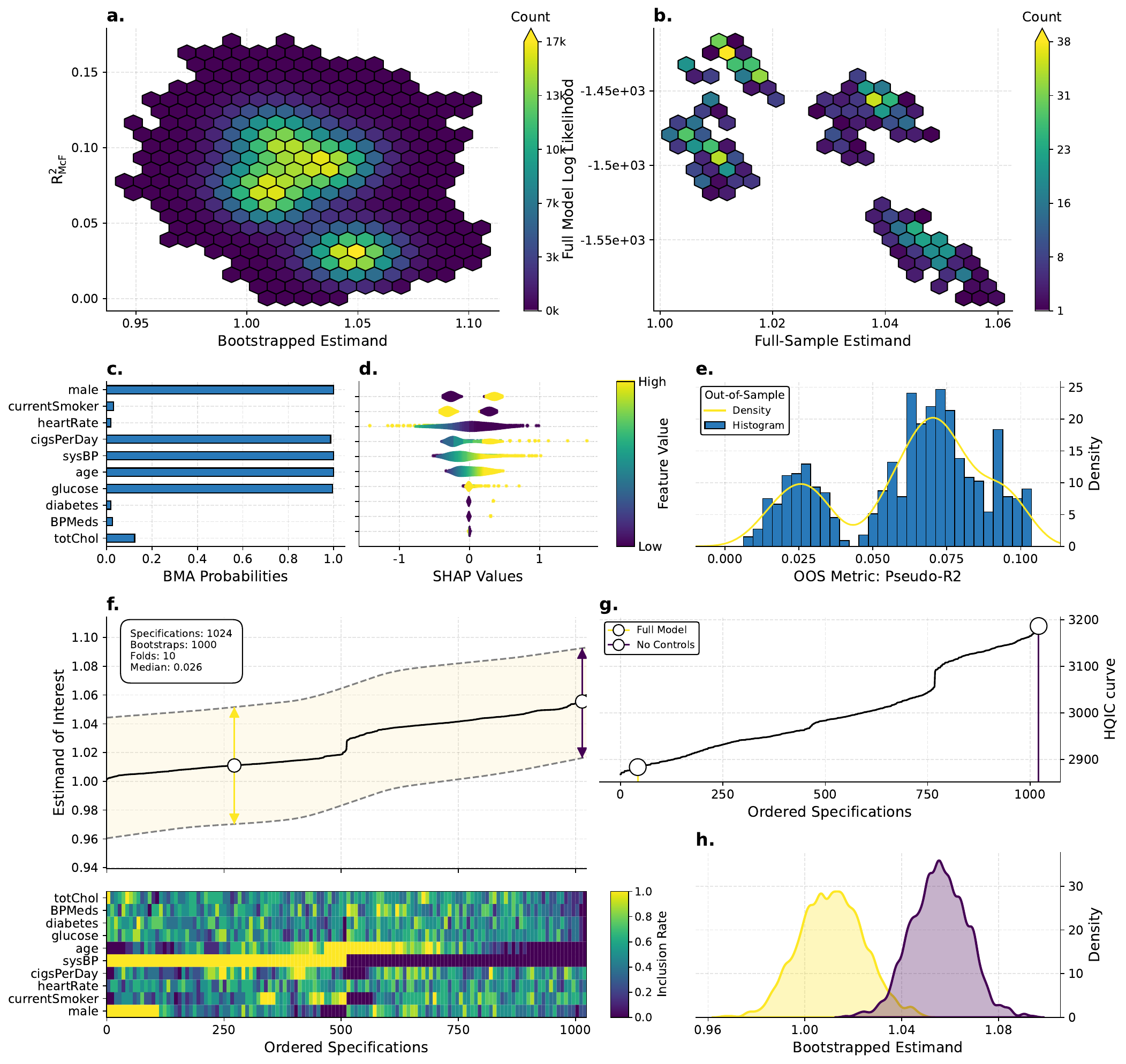}
	\caption{\textbf{Example output of \texttt{RobustiPy} with a binary dependent variable}. The dependent variable is a ten year metric of coronary heart disease (CHD), and the variable of interest is Body Mass Index.}
	\label{figure:framingham}
\end{figure}

\clearpage

\begin{figure}[ht]
	\centering
	\includegraphics[width=\textwidth]{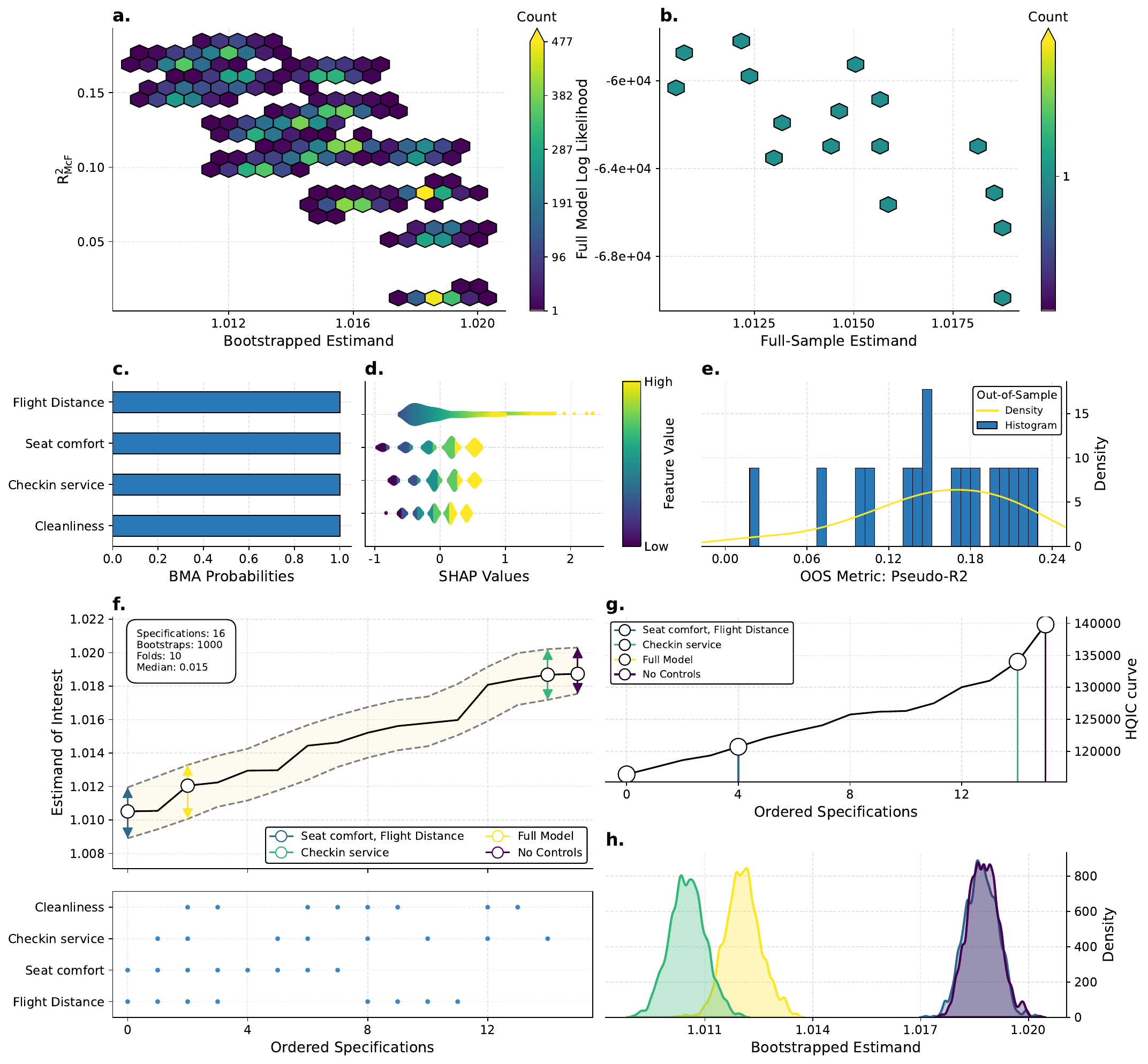}
	\caption{\textbf{Example output of \texttt{RobustiPy} utilising a binary dependent variable}. The dependent variable is a representation of custom satisfaction within an anonymous airline company, with data provided by Kaggle. The estimand of interest relates to the age of the customer.}
	\label{fig:S_airline}
\end{figure}

\newpage

\subsection*{Supplementary Code Blocks}

\begin{suppcodeblock}{Array of never-changing predictors}{scb:never-changing}
	\begin{minted}[baselinestretch=1,fontsize=\footnotesize]{python}
import numpy as np
import pandas as pd
from robustipy.models import OLSRobust


def sim2(project_name):
    np.random.seed(192735)
    beta1 = np.array([0.2, 0.5, -0.4, -0.7, 0.2,
                      0.5, 0.2, 0.5, 0.3])
    L = np.array([
        [0.8, 0.2], [0.6, -0.5], [0.7, 0.1],
        [0.5, -0.6], [0.4, 0.7], [0.3, -0.4],
        [0.2, 0.3], [0.1, -0.2]])
    D = np.diag([5] * 8)
    cov_matrix = L.dot(L.T) + D
    num_samples = 1000
    mean_vector = np.zeros(8)
    X = np.random.multivariate_normal(mean=mean_vector,
                                      cov=cov_matrix,
                                      size=num_samples)
    X_i = np.column_stack((np.ones(num_samples), X))
    errors = np.random.normal(0.0, 1.0, num_samples)
    Y1 = np.dot(X_i, beta1) + errors
    np_data = np.column_stack((Y1, X))
    data = pd.DataFrame(np_data, columns=['y1', 'x1', 'z1', 'z2',
                                          'z3', 'z4', 'z5', 'z6',
                                          'z7'])
    y = ['y1']
    x = ['x1', 'z1']
    z = ['z2', 'z3','z4', 'z5', 'z6', 'z7']
    sim2 = OLSRobust(y=y, x=x, data=data)
    sim2.fit(controls=z, draws=1000, kfold=10, seed=192735)
    sim2_results = sim2.get_results()
    sim2_results.plot(specs=[['z4', 'z5']],
                      ic='hqic', figsize=(16, 16),
                      ext='pdf',
                      project_name=project_name)
    sim2_results.summary()


if __name__ == "__main__":
    sim2('sim2_example')
\end{minted}
\end{suppcodeblock}

\newpage

\begin{suppcodeblock}{Fixed Effect OLS}{scb:fixed}
	\begin{minted}[baselinestretch=1,fontsize=\footnotesize]{python}
import numpy as np
import pandas as pd
from robustipy.models import OLSRobust


def sim3(project_name):
    np.random.seed(192735)
    beta_mean = [0.9, 0.4, -0.7, -0.1, 0.2, 0.3, -0.2, 0.5, 0.1]
    beta_std = 2
    L_factor = np.array([[0.8,  0.2], [0.6, -0.5],
                         [0.7,  0.1], [0.5, -0.6],
                         [0.4,  0.7], [0.3, -0.4],
                         [0.2,  0.3], [0.1, -0.2]])
    D = np.diag([0.3] * 8)
    cov_matrix = L_factor.dot(L_factor.T) + D
    num_samples, num_groups = 10, 1000
    mean_vec = np.zeros(8)
    X_list, Y_list, group_list = [], [], []
    for group in range(1, num_groups + 1):
        beta = np.random.normal(beta_mean, beta_std, len(beta_mean))
        mi, ma, factor = np.argmin(beta), np.argmax(beta), 5
        beta[mi] = np.random.normal(beta[mi], beta_std * factor)
        beta[ma] = np.random.normal(beta[ma], beta_std * factor)
        X = np.random.multivariate_normal(mean_vec,
                                          cov_matrix,
                                          num_samples) \
            + np.random.normal(0, 5, (num_samples, 8))
        X_i = np.column_stack((np.ones(num_samples), X))
        errors = np.random.normal(0, 20, num_samples)
        Y = np.dot(X_i, beta) + errors
        X_list.append(X)
        Y_list.append(Y)
        group_list.extend([group] * num_samples)
    data = pd.DataFrame(np.column_stack((np.concatenate(Y_list),
                                         np.vstack(X_list))),
                        columns=['y', 'x1', 'z1', 'z2', 'z3',
                                 'z4', 'z5', 'z6', 'z7'])
    data['group'] = group_list
    sim3 = OLSRobust(y=['y'], x=['x1'], data=data)
    sim3.fit(controls=['z1', 'z2', 'z3', 'z4', 'z5', 'z6', 'z7'],
             group='group', draws=1000, kfold=10,
             rescale_y=True, rescale_z=True,
             seed=192735)
    sim3_results = sim3.get_results()
    sim3_results.plot(specs=[['z1', 'z2', 'z3']], ic='hqic',
                      figsize=(16, 16), ext='pdf',
                      project_name=project_name)
    sim3_results.summary()


if __name__ == '__main__':
    sim3('sim3_example')
\end{minted}
\end{suppcodeblock}

\newpage

\begin{suppcodeblock}{Logistic Regression}{scb:logisticregression}
	\begin{minted}[baselinestretch=1,fontsize=\footnotesize]{python}
import numpy as np
import pandas as pd
from robustipy.models import LRobust

def sim4(project_name):
    np.random.seed(192735)
    beta1 = np.array([0.05, 0.1, -0.6, -0.35, 0.05,
                      0.1, 0.05, 0.1, 0.05])
    L_factor = np.array([
        [0.8,  0.2],
        [0.6, -0.5],
        [0.7,  0.1],
        [0.5, -0.6],
        [0.4,  0.7],
        [0.3, -0.4],
        [0.2,  0.3],
        [0.1, -0.2]
    ])
    D = np.diag([0.3]*8)
    cov_matrix = L_factor.dot(L_factor.T) + D
    num_samples = 10000
    mean_vector = np.zeros(8)
    X = np.random.multivariate_normal(mean=mean_vector,
                                      cov=cov_matrix,
                                      size=num_samples)
    X_i = np.column_stack((np.ones(num_samples), X))
    errors = np.random.normal(0.0, 1, num_samples)
    Y1 = np.dot(X_i, beta1) + errors
    threshold = np.median(Y1)
    Y1 = (Y1 > threshold).astype(int)
    np_data = np.column_stack((Y1, X))
    data = pd.DataFrame(np_data, columns=['y1','x1','z1','z2',
                                          'z3','z4','z5','z6','z7'])

    y = ['y1']
    x = ['x1']
    z = ['z1','z2','z3','z4','z5','z6','z7']
    sim4 = LRobust(y=y, x=x, data=data)
    sim4.fit(controls=z, draws=1000, kfold=10, seed=192735)
    sim4_results = sim4.get_results()
    sim4_results.plot(specs=[['z1'], ['z2','z4'],
                             ['z3','z5']],
                      ic='hqic', figsize=(16,16),
                      ext='pdf', project_name=project_name)
    sim4_results.summary()


if __name__ == "__main__":
    sim4('sim4_example')
\end{minted}
\end{suppcodeblock}

\end{document}